\documentclass[aps,prd,onecolumn,eqsecnum,amsmath,nofootinbib,preprintnumbers,superscriptaddress]{revtex4-2}

\usepackage[dvipsnames]{xcolor}
\usepackage{color,graphicx,float,xcolor}
\usepackage{amsfonts,amssymb,theorem,mathrsfs,times}
\usepackage{bm}
\usepackage{multirow}
\usepackage{mathtools}
\usepackage{amsfonts,amssymb,theorem,mathrsfs}
\usepackage{dsfont}
\usepackage{setspace}
\usepackage{amsmath}
\usepackage{graphicx}
\usepackage[colorlinks,linkcolor=blue,anchorcolor=blue,citecolor=blue]{hyperref}
\usepackage{ulem}
\usepackage{cancel}
\usepackage[justification = raggedright, singlelinecheck = true]{caption}
\usepackage{subcaption}
\usepackage{tikz}
\usetikzlibrary{decorations.pathmorphing,calc,arrows.meta,intersections}

\begin{document}
\title{The entropy of dynamical black holes deviating from electrovacuum in second order}
\preprint{\hfill {\small {ICTS-USTC/PCFT-26-45}}}
\date{\today}

\author{Wen-Tao Fu}
\email{fuwentao2024@mail.ustc.edu.cn}

\affiliation{Interdisciplinary Center for Theoretical Study and Department of Modern Physics,\\
University of Science and Technology of China, Hefei, Anhui 230026, China}

\author{Ming-Fei Ji}
\email{jimingfei@mail.ustc.edu.cn}

\affiliation{Interdisciplinary Center for Theoretical Study and Department of Modern Physics,\\
University of Science and Technology of China, Hefei, Anhui 230026, China}

\author{Yu-Sen Zhou}
\email{zhou\_ys@mail.ustc.edu.cn}

\affiliation{Interdisciplinary Center for Theoretical Study and Department of Modern Physics,\\
University of Science and Technology of China, Hefei, Anhui 230026, China}

\author{Li-Ming Cao}
\email{caolm@ustc.edu.cn}
\affiliation{Interdisciplinary Center for Theoretical Study and Department of Modern Physics,\\
University of Science and Technology of China, Hefei, Anhui 230026, China}
\affiliation{Peng Huanwu Center for Fundamental Theory, Hefei, Anhui 230026, China}

\begin{abstract}
We study the entropy of dynamical black holes in Einstein--Maxwell theory. Starting from a stationary electrovacuum background with a bifurcate Killing horizon, the apparent horizon is constructed perturbatively in Gaussian null coordinates, and the area of its cross section is derived to second order. Within the covariant phase space formalism, we introduce a modified canonical energy that incorporates contributions from external matter and is naturally related to the second-order variation of the entropy through the balance law. Under the null energy condition and the asymptotic stationary condition, the entropy is proportional to the area of the apparent horizon up to second order and satisfies the second law. Without the null energy condition, establishing this area law requires additional assumptions.
\end{abstract}

\maketitle

\section{Introduction}
\label{sec:introduction}

A central question in black-hole physics is how thermodynamic concepts are connected to the spacetime geometry and the underlying theory of gravity. This connection is most clearly established in the stationary case, as illustrated by stationary black holes in general relativity, whose event horizons provide a natural geometric setting and have nondecreasing area under the null energy condition. This analogy to the second law of thermodynamics, together with the other laws of black-hole mechanics, motivated the identification of the entropy of black hole with the horizon area~\cite{Bekenstein:1973ur,Bardeen:1973gs}. Hawking's discovery of black-hole evaporation subsequently led to the Bekenstein--Hawking formula $S=A/(4G\hbar)$~\cite{Hawking:1975vcx}. In this way, thermodynamics of stationary black holes provides a direct link between classical gravity, quantum field theory, and statistical physics~\cite{Wald:1999vt}.

Later developments in the covariant phase space formalism provided a more systematic understanding of the entropy of black hole. In this framework, the entropy of a stationary black hole is identified with the Noether charge associated with the horizon-generating Killing field~\cite{Lee:1990nz,Wald:1990mme,Wald:1993nt,Jacobson:1993vj,Iyer:1994ys}. This construction is particularly powerful because it applies to general diffeomorphism covariant theories and does not rely on a specific theory of gravity. However, the standard derivation relies essentially on stationarity and the existence of a horizon-generating Killing field. Once the black hole is driven away from equilibrium, no preferred Killing symmetry is available, and it becomes nontrivial to determine which geometric surface and which entropy functional should characterize its thermodynamics.

This issue is closely related to the distinction between event horizons and apparent horizons. The event horizon is defined globally and depends on the future null infinity of spacetime, whereas the apparent horizon is characterized quasi-locally by marginally trapped surfaces on a chosen spacelike foliation (even without this foliation~\cite{Ashtekar_2004}). For dynamical processes, it is therefore natural to ask whether entropy should be associated with cross sections of the event horizon or with such marginally trapped surfaces. A recent construction by Wald and collaborators, based on the covariant phase space formalism, argued that at first order, the dynamical entropy is naturally associated with the apparent horizon rather than the event horizon~\cite{Hollands:2024vbe}. Related developments involving higher-curvature entropy functionals and the linearized second law likewise indicate that an appropriate entropy functional away from equilibrium may differ from the standard Iyer--Wald entropy expression for stationary black holes evaluated on an arbitrary event horizon slice~\cite{Dong:2013qoa,Wall:2015raa,Rignon-Bret:2023fjq,Furugori:2025pmn}.

Beyond linear order, the analysis becomes more subtle because nonlinear effects, interactions between different perturbative contributions, gauge choices, and ambiguities in the relevant quantities must all be taken into account. Second-order perturbations in vacuum have been studied in several contexts~\cite{Hollands:2024vbe,Kong:2024sqc}, and our previous work derived an explicit second-order entropy formula involving only the gravitational contribution~\cite{Fu:2026itf}. In Einstein--Maxwell theory, the balance law acquires additional electromagnetic fluxes through future event horizon and future null infinity. These contributions must be included to determine whether the entropy remains proportional to the area of the apparent horizon at second order.

An alternative quasi-local viewpoint on nonequilibrium black-hole thermodynamics has also been developed by Ashtekar et al. Recently, they proposed a nonperturbative framework in which entropy is associated with marginally trapped cross sections of quasi-local horizons far from equilibrium~\cite{Ashtekar:2025qqa,Ashtekar:2026jdz}. This approach is conceptually distinct from the covariant phase space construction, since it is based on quasi-local horizon mechanics and usual canonical phase space analysis~\cite{Ashtekar_2004}. 

Nevertheless, both viewpoints indicate that marginally trapped surfaces are central to the thermodynamic description of dynamical black holes but neither incorporates the Maxwell field. The Maxwell field is not merely another matter field but a particularly natural and fundamental extension of vacuum general relativity. As a long-range gauge field, it carries a conserved charge measurable at infinity, contributes directly to black hole thermodynamics, and gives rise to the Kerr–Newman family, the standard family of stationary charged black holes. Under the usual assumptions of black-hole uniqueness theorems, this family exhausts the stationary, asymptotically flat electrovacuum black-hole solutions~\cite{Chrusciel:2012jk}. It also plays a central role in black-hole thermodynamics, with the electric charge and its conjugate potential entering explicitly into the first law~\cite{Gao:2003ys}. A natural extension of the thermodynamic analysis from stationary to dynamical black holes should therefore incorporate the Maxwell field. This raises the question of how electric charge modifies the area law in the dynamical regime. It remains unclear up to date how the Maxwell field can be incorporated into the approach of Ashtekar et al., while it has been incorporated at first order within the covariant phase space formalism~\cite{Visser:2025jnf}. This provides a natural starting point for investigating whether the relation between entropy and the area of apparent horizon extends to second order in Einstein--Maxwell theory.

In this work, we investigate the entropy of black holes deviating from stationary electrovacuum black holes at second order in Einstein--Maxwell theory. Working in Gaussian null coordinates, we construct the apparent horizon perturbatively near the event horizon and obtain the corresponding expansion relations and second-order area formula. These geometric results are then combined with the covariant phase space formalism, and a modified canonical energy adapted to perturbations sourced by external matter is introduced. A relation between the modified canonical energy and entropy is established through the balance law. We show that, under the null energy condition and the asymptotic stationary condition, the second-order dynamical entropy is proportional to the area of the apparent horizon. Without the null energy condition, this proportionality is not automatic and requires additional conditions. Our result extends the proposal of~\cite{Hollands:2024vbe} to second order in Einstein--Maxwell theory and supports the interpretation of the apparent horizon as the relevant geometric characterization of the entropy of black hole. The geometric construction and the part of the canonical energy analysis involving only the gravitational contribution follow~\cite{Fu:2026itf}.

The rest of this paper is organized as follows. In Section~\ref{sec:geometry_and_horizon}, we introduce the near-horizon geometry in Gaussian null coordinates, specify the perturbative gauge conditions, and describe the perturbative construction of the apparent horizon from the vanishing-expansion condition. We then summarize the expansion relations and the second-order area formula. In Section~\ref{sec:modified_canonical_energy}, we review the symplectic-current identity in the covariant phase space formalism, define the modified canonical energy, and evaluate the fluxes through $\mathcal H^+$ and $\mathcal I^+$, including the electromagnetic contributions in Einstein--Maxwell theory. In Section~\ref{sec:dynamical_entropy}, we express the balance law in terms of modified canonical energy forms, obtain the second-order entropy formula, and compare it with the area of the apparent horizon. We also discuss the role of the null energy condition in establishing the area law. A summary and discussion are provided in Section~\ref{sec:conclusion}.

\section{Black Hole Geometry and Apparent Horizon}
\label{sec:geometry_and_horizon}

In this section, we introduce the near-horizon geometry in Gaussian null coordinates and specify the gauge conditions used in the perturbative analysis. We then construct the apparent horizon perturbatively near the event horizon and summarize the expansion relations and second-order area formula needed for the comparison with the dynamical entropy.

We begin with a stationary black-hole background $(M,g)$ in $n$-dimensional spacetime. The background event horizon is assumed to be a bifurcate Killing horizon $\mathcal H$, with future branch $\mathcal H^+$, past branch $\mathcal H^-$, and bifurcation surface $\mathcal B$~\cite{Visser:2024pwz}. In the neighborhood of $\mathcal H^+$, we introduce Gaussian null coordinates $(u,v,x^i)$, where $i=1,\ldots,n-2$. The future horizon is located at $u=0$, and $v$ is an affine parameter along the null generators. The metric is written as~\cite{Furugori:2025pmn,Jia:2025tgf,Visser:2024pwz}
\begin{equation}
  ds^2 = -2du\,dv - u^2\alpha \,dv^2 - 2u\beta_i \,dv\,dx^i + \gamma_{ij} \,dx^i\,dx^j\, .
  \label{eq:main_gnc_metric_expanded}
\end{equation}
Here $\alpha$, $\beta_i$, and $\gamma_{ij}$ are functions of $(u,v,x^i)$, and $\gamma_{ab}=\gamma_{ij} (dx^i)_a(dx^j)_b$ is the induced metric on the codimension-two cross sections. The Gaussian null coordinates in metric retain a residual coordinate freedom, as discussed in Appendix~\ref{app:residual_gnc_freedom}.

The basis vectors $k^a=(\partial_v)^a$, $l^a=(\partial_u)^a$, and $m^a_i=(\partial_i)^a$ satisfy
\begin{equation}
  k_a = g_{ab}k^b = -(du)_a -u^2\alpha(dv)_a -u\beta_i(dx^i)_a \overset{\mathcal H^+}{=} -(du)_a\, , \qquad l_a = g_{ab}l^b = -(dv)_a\, .
\label{eq:main_null_covectors_expanded}
\end{equation}
It follows that $k_al^a=-1$, $l_al^a=0$, $k_ak^a\overset{\mathcal H^+}{=}0$, and the spacetime metric has a decomposition $g_{ab}\overset{\mathcal H^+}{=}-2k_{(a}l_{b)}+\gamma_{ab}$. The horizon-generating Killing field is normalized as~\cite{Visser:2024pwz}
\begin{equation}
  \xi^a=\kappa(vk^a-ul^a)\, ,
\label{eq:main_xi_expanded}
\end{equation}
where $\kappa$ is the surface gravity. This form makes explicit that $\xi^a=0$ on the bifurcation surface and that $\xi^a=\kappa vk^a$ on $\mathcal H^+$. Stationarity implies that the metric functions depend on $u$ and $v$ through the boost-invariant combination $uv$~\cite{Hollands:2022fkn,Bhattacharyya:2021jhr}.

Let $g_{ab}(s)$ be a one-parameter family of metrics expanded as 
\begin{equation}
  g_{ab}(s)=g_{ab}+s\delta g_{ab}+\frac{s^2}{2}\delta^2g_{ab}+O(s^3)\, ,
\label{eq:main_metric_family_expanded}
\end{equation}
where $s=0$ corresponds to the stationary background. The diffeomorphism gauge freedom is used to identify the perturbed event horizon with the background horizon. This leads to the same gauge conditions adopted in~\cite{Fu:2026itf}. In particular, $\mathcal H^+$ and  $\mathcal H^-$ remain at $u=0$ and  $v=0$ respectively, and the vector fields $k^a$ and $l^a$ are held fixed under the perturbation:
\begin{equation}
  \delta^m k^a=0\, , \qquad \delta^m l^a=0\, , \qquad (m=1,2)\, .
\label{eq:main_kl_fixed_expanded}
\end{equation}
The null and normalization conditions are also imposed in the perturbed spacetime, which imply that
\begin{equation}
  k^a\delta^m g_{ab}\overset{\mathcal H^+}{=}0\, , \qquad l^a\delta^m g_{ab}=0\, , \qquad (m=1,2)\, .
\label{eq:main_gauge_metric_expanded}
\end{equation}
It is also required that $k^a$ remains affinely parametrized on $\mathcal H^+$ and that $l^a$ remains affinely parametrized throughout the spacetime under perturbations. These conditions are the geometric gauge choices used throughout the computation and ensure that perturbations are represented by the changes of the functions $\alpha$, $\beta_i$, and $\gamma_{ij}$. A subtle point is that $\xi^a$ is not held fixed under perturbations. Under the perturbation, $\xi^a$ is required to remain null and tangent to the generators of the perturbed horizon. Together with Eq.~\eqref{eq:main_gauge_metric_expanded}, this implies that on $\mathcal H^+$ the variation $\delta^m\xi^a$ is proportional to $k^a$, while on $\mathcal H^-$ it is proportional to $l^a$.

Roughly speaking, the apparent horizon can be understood as the envelope of the marginally trapped surfaces, which deviates from the event horizon, $u=0$, under perturbations. The position of the apparent horizon $\mathcal A$ is denoted by~\cite{Visser:2024pwz}
\begin{equation}
 u=U(s;v,x) =s\delta U(v,x)+\frac{s^2}{2}\delta^2U(v,x)+O(s^3)\, . 
\label{eq:main_apparent_horizon_location}
\end{equation}
We assume that $U(s;v,x)$ is smooth in $s$ and that $U(0;v,x)=0$. For a fixed $v$, let $\mathcal T(v)$ denote the cross section of $\mathcal A$, with the ingoing null normal chosen as $\tilde l=-(dv)$, and the outgoing null normal $\tilde k_a$ constructed as a linear combination of the two normal one-forms $d(u-U(s;v,x))$ and $dv$. The coefficients are fixed by the normalization conditions $\tilde k^a\tilde l_a=-1$ and $\tilde k^a\tilde k_a=0$ up to $\mathcal{O}(s^2)$~\cite{Fu:2026itf}. On $\mathcal{T}(v)$, $U$ is determined by requiring the outgoing expansion to vanish, i.e., $\tilde{\theta}_{\tilde{k}}=0$, order by order. At zeroth order, one recovers the stationary condition $\theta_v=0$. At first order, we use the relation $\theta_u(0,v,x)=-v\mu(x)$ from~\cite{Wall:2015raa,Bhattacharyya:2021jhr,Hollands:2022fkn,Wall:2024lbd,Jia:2025tgf}, assuming $\mu(x)>0$ for the chosen cross sections, to obtain
\begin{equation}
\delta\theta_v=\mu\delta U-D(D\delta U+\beta\delta U)\, .
\label{eq:main_delta_theta_relation_expanded}
\end{equation}
Here $D_i$ is the intrinsic covariant derivative on the horizon cross section, and we have suppressed the indices on $D_i$, $\beta_i$, and $\gamma_{ij}$ in the same way as in the detailed calculation. At second order, the resulting relation is more complicated. Schematically, it takes the form
\begin{equation}
\begin{aligned}
\delta^2\theta_v ={}&\mu\delta^2U-(\delta U)^2\partial_u^2\theta_v -2\delta U\partial_u\delta\theta_v-(D\delta U)^2\theta_u-\frac{\lambda v}{\mu} \left[\delta\theta_v+D_i(D^i\delta U+\beta^i\delta U)\right]^2\\
&-D_i\Bigl[D^i\delta^2U+\beta^i\delta^2U +2(\delta\gamma^{ij}D_j\delta U+\delta\beta^i\delta U)+2\delta U(\partial_u\gamma^{ij}D_j\delta U +\delta U\partial_u\beta^i)\Bigr]\\
&-(D^i\delta U+\beta^i\delta U)D_i(\gamma^{jk}\delta\gamma_{jk}) -2\delta U(D^i\delta U+\beta^i\delta U)D_i\theta_u\, .
\end{aligned}
\label{eq:main_delta2_theta_relation_expanded}
\end{equation}
where $\lambda=\alpha+\beta^2$.

We next expand the area of $\mathcal T(v)$. Let $dA_\epsilon(0,v,x)$ be the area element of the corresponding event-horizon cross section $\mathcal C(v)$. Using $\partial_vdA_\epsilon=\theta_vdA_\epsilon$ and $\partial_udA_\epsilon=\theta_udA_\epsilon$, the expansion in $U$ then rewrites the area of the apparent horizon in terms of quantities on $\mathcal H^+$. By using Eqs.~\eqref{eq:main_delta_theta_relation_expanded} and~\eqref{eq:main_delta2_theta_relation_expanded}, and dropping the total derivative terms on the compact cross sections, the second-order variation of the area is
\begin{equation}
\begin{aligned}
\delta^2A_{\mathcal T}(v) ={}&\delta^2\int_{\mathcal C}dA_\epsilon(1-v\theta_v) +\int_{\mathcal C}dA_\epsilon\Biggl\{ \left(1+\frac{\lambda v}{\theta_u}\right)(\theta_u\delta U)^2 -v(D\delta U)^2\theta_u+2\delta U(\delta\theta_u-v\partial_u\delta\theta_v)\\
&\quad+(\delta U)^2(\partial_u\theta_u-v\partial_u^2\theta_v)-2v\delta U(D^i\delta U+\beta^i\delta U)D_i\theta_u \Biggr\}\, .
\end{aligned}
\label{eq:main_area_expansion_expanded}
\end{equation}
More details can be found in~\cite{Fu:2026itf}.

\section{Modified Canonical Energy}
\label{sec:modified_canonical_energy}
In this section, we review the symplectic-current identity and introduce a modified canonical energy in the presence of external matter sources. The external matter is assumed to carry no electromagnetic current and therefore does not source the Maxwell field. We then evaluate the fluxes through the horizon and future null infinity, including the Einstein--Maxwell contributions, and absorb the boundary terms into $\mathcal E'[\Sigma(t)]$.

The symplectic-current identity used below has been derived in~\cite{Fu:2026itf}. In particular, for perturbations around a stationary background satisfying $\mathcal L_\xi\phi=0$, one obtains the symplectic-current identity~\cite{Hollands:2024vbe}
\begin{equation}
\boldsymbol\omega(\phi;\delta\phi,\mathcal L_\xi\delta\phi) = \xi\cdot[\delta\boldsymbol E(\phi)\delta\phi] +\xi\cdot[\boldsymbol E(\phi)\delta^2\phi] +\xi^a\delta^2\boldsymbol C_a(\phi) +d\left[ \delta_\phi^2\boldsymbol Q_\xi(\phi) -\xi\cdot\delta\boldsymbol\Theta(\phi,\delta\phi) \right]\, ,
\label{eq:main_second_noether_expanded}
\end{equation}
where the complete set of dynamical fields is represented by $\phi$, and $\delta_\phi$ denotes variation with respect to $\phi$ alone, with $\xi^a$ held fixed. The equations of motion are given by $\boldsymbol E(\phi)=0$, while the symplectic potential $(n-1)$-form and the Noether-charge $(n-2)$-form associated with the vector field $\xi^a$ are denoted by $\boldsymbol\Theta(\phi,\delta\phi)$ and $\boldsymbol Q_{\xi}(\phi)$, respectively~\cite{Wald:1993nt,Iyer:1994ys}. In the above equation, $\boldsymbol C_a(\phi)$ is the dual vector-valued $(n-1)$-form which vanishes on shell and is given by~\cite{Hollands:2024vbe}
\begin{eqnarray}
  \boldsymbol C_a(\phi) &=&\boldsymbol\epsilon_c^{(n)} \Bigg[ 2(E_G)^{c}{}_{a} - \sum_A A^{d_1\cdots d_k}{}_{b_1\cdots a \cdots b_l}\, (E_M)_{d_1\cdots d_k}{}^{b_1\cdots c \cdots b_l} \nonumber\\
    &&+ \sum_A A^{d_1\cdots c \cdots d_k}{}_{b_1\cdots b_l}\, (E_M)_{d_1\cdots a \cdots d_k}{}^{b_1\cdots b_l} \Bigg] \, .
  \label{eq:full_Ca_expression}
\end{eqnarray}
Here, $\boldsymbol\epsilon_c^{(n)}=\epsilon_{c a_1\cdots a_{n-1}}$ is the spacetime volume form, $(E_G)_{ab}=0$ are the equations of motion for $g_{ab}$, and $(E_M)_{a_1\cdots a_k}{}^{b_1\cdots b_l}=0$ are the equations of motion for the matter field $A^{a_1\cdots a_k}{}_{b_1\cdots b_l}$.

Following~\cite{Fu:2026itf}, we define the ``canonical energy form'' by
\begin{equation}
\boldsymbol\omega'(\phi;\delta\phi,\mathcal L_\xi\delta\phi) = \boldsymbol\omega(\phi;\delta\phi,\mathcal L_\xi\delta\phi) -\xi\cdot[\delta\boldsymbol E(\phi)\delta\phi] -\xi\cdot[\boldsymbol E(\phi)\delta^2\phi] -\xi^a\delta^2\boldsymbol C_a(\phi)\, ,
\label{eq:main_modified_current_expanded}
\end{equation}
which is closed according to Eq.~\eqref{eq:main_second_noether_expanded}. The corresponding canonical energy on a Cauchy surface $\Sigma(t)$ is
\begin{equation}
  \mathcal E[\Sigma(t)] = \int_{\Sigma(t)} \boldsymbol\omega'(\phi;\delta\phi,\mathcal L_\xi\delta\phi)\, .
\label{eq:main_modified_energy_expanded}
\end{equation}
Applying Stokes' theorem to the compact region bounded by $\Sigma(t_1)$, $\Sigma(t_2)$, $\mathcal H^+_{12}$, and $\mathcal I^+_{12}$ gives
\begin{equation}
  \mathcal E[\Sigma(t_2)]-\mathcal E[\Sigma(t_1)] = -\int_{\mathcal H^+_{12}}\boldsymbol\omega' -\int_{\mathcal I^+_{12}}\hat{\boldsymbol\omega}'\, ,
\label{eq:main_canonical_balance_expanded}
\end{equation}
where $\Sigma(t_1)$ and $\Sigma(t_2)$ are Cauchy surfaces with $t_1<t_2$, and the portions of the event horizon $\mathcal H^+$ and future null infinity $\mathcal I^+$ bounded by their intersections with these surfaces are denoted by $\mathcal H^+_{12}$ and $\mathcal I^+_{12}$, respectively.

In the following analysis, we specialize to four-dimensional  Einstein--Maxwell theory with $\phi=(g,A)$, and split the symplectic current into gravitational and electromagnetic pieces~\cite{Sorce:2017dst}
\begin{equation}
  \boldsymbol\omega(\phi;\delta\phi,\mathcal L_\xi\delta\phi) = \boldsymbol\omega^{GR}(g;\delta g,\mathcal L_\xi\delta g) + \boldsymbol\omega^{EM}(\phi;\delta\phi,\mathcal L_\xi\delta\phi)\, .
\label{eq:main_split_gr_em_expanded}
\end{equation}
The detailed derivations of the flux formulas below are collected in Appendix~\ref{app:modified_canonical_energy}. For the gravitational contribution to the symplectic-current flux through $\mathcal H^+_{12}$, integrating the first term of Eq.~\eqref{eq:main_split_gr_em_expanded}, we have
\begin{equation}
\begin{aligned}
  \int_{\mathcal{H}^+_{12}}\boldsymbol\omega^{GR} \overset{\mathcal{H}^+}{=}& \frac{1}{4\pi G} \int_{\mathcal H^+_{12}} \boldsymbol\epsilon^{(3)} (\xi^c\nabla_c v) \left[ \delta\sigma^{ab}\delta\sigma_{ab} -\frac12(\delta\theta_v)^2 \right]\\
  &+ \frac{1}{32\pi G} \Delta \int_{\mathcal{C}} \boldsymbol {\epsilon}^{(2)} (\xi^c\nabla_c v) \left( \delta \gamma^{ab}\partial_v \delta \gamma_{ab} + 2\delta \theta_v \gamma^{ab}\delta \gamma_{ab} \right)\, ,
\label{eq:main_gr_horizon_flux_expanded}
\end{aligned}
\end{equation}
where $\Delta$ denotes the difference of a quantity evaluated on the two boundary cross sections bounding $\mathcal H^+_{12}$, $\boldsymbol\epsilon^{(3)}$ is the volume $3$-form on $\mathcal H^+$, and $\boldsymbol\epsilon^{(2)}$ is the area $2$-form on $\mathcal C$. For the electromagnetic contribution to the symplectic-current flux through $\mathcal H^+_{12}$, integrating the second term of Eq.~\eqref{eq:main_split_gr_em_expanded}, we find
\begin{equation}
  \int_{\mathcal{H}^+_{12}}\boldsymbol\omega^{EM} \overset{\mathcal H^+}{=} -\frac{1}{4\pi G} \Delta\int_{\mathcal{C}} \boldsymbol{\epsilon}^{(2)} (\xi^c\nabla_c v) k_a\delta A_b\delta F^{ab} + \int_{\mathcal H^+_{12}} \boldsymbol{\epsilon}^{(3)} (\xi^c\nabla_c v) k^a k^b\delta^2 T^{EM}_{ab}\, .
\label{eq:main_em_horizon_flux_expanded}
\end{equation}
The contribution of the constraint is
\begin{equation}
  \begin{aligned}
  -\int_{\mathcal H^+_{12}}\xi^a\delta^2\boldsymbol C_a \overset{\mathcal H^+}{=} \int_{\mathcal H^+_{12}} (\xi^c\nabla_c v) \left[ \delta\boldsymbol{\epsilon}^{(3)} 2k^a k^b\delta T_{ab} + \boldsymbol{\epsilon}^{(3)} k^a k^b\delta^2T_{ab} \right]\, .
  \end{aligned}
\label{eq:main_constraint_horizon_flux_expanded}
\end{equation}
Here, $\delta T_{ab}$ and $\delta^2T_{ab}$ are the first- and second-order stress-energy perturbations of the external matter, determined by the metric equations of motion encoded in $\boldsymbol C_a(\phi)$. Combining Eqs.~\eqref{eq:main_gr_horizon_flux_expanded}, \eqref{eq:main_em_horizon_flux_expanded} and \eqref{eq:main_constraint_horizon_flux_expanded}, the contribution of Eq.~\eqref{eq:main_modified_current_expanded} on $\mathcal{H}^+$ becomes
\begin{equation}
\begin{aligned}
  \int_{\mathcal{H}^+_{12}}\boldsymbol\omega' \overset{\mathcal H^+}{=}\;& \frac{1}{4\pi G} \int_{\mathcal H^+_{12}} \boldsymbol\epsilon^{(3)} (\xi^c\nabla_c v) \left[ \delta\sigma^{ab}\delta\sigma_{ab} -\frac12(\delta\theta_v)^2 \right]  \\
  &+ \frac{1}{32\pi G} \Delta \int_{\mathcal{C}} \boldsymbol {\epsilon}^{(2)} (\xi^c\nabla_c v) \left( \delta \gamma^{ab}\partial_v \delta \gamma_{ab} + 2\delta \theta_v \gamma^{ab}\delta \gamma_{ab} \right) \\
  &- \frac{1}{4\pi G} \Delta\int_{\mathcal{C}} \boldsymbol{\epsilon}^{(2)} (\xi^c\nabla_c v) k_a\delta A_b\delta F^{ab}  \\
  &+ \int_{\mathcal H^+_{12}} (\xi^c\nabla_c v) \left[ \delta\boldsymbol{\epsilon}^{(3)} 2k^a k^b\delta T'_{ab} + \boldsymbol{\epsilon}^{(3)} k^a k^b\delta^2T'_{ab} \right]\, .
\end{aligned}
\label{eq:main_omega_prime_horizon_flux_expanded}
\end{equation}
In this expression, $T'_{ab}$ denotes the total stress-energy tensor, including contributions from both external matter and the electromagnetic field.

We now turn to future null infinity $\mathcal I^+$. To evaluate the contribution of $\boldsymbol\omega'(\phi;\delta\phi,\mathcal L_\xi\delta\phi)$ on $\mathcal I^+_{12}$, an unphysical spacetime is introduced by $\hat g_{ab}=\Omega^2 g_{ab}$, where $\Omega$ is the conformal factor. Near $\mathcal I^+$, a Bondi frame is chosen in which the unphysical metric takes the asymptotic form~\cite{Hollands:2012sf}
\begin{equation}
  \hat g_{ab} = (d\Omega)_a(d\hat u)_b + (d\Omega)_b(d\hat u)_a + \hat\gamma_{ij}(dx^i)_a(dx^j)_b + \mathcal O(\Omega)\, .
\label{eq:main_bondi_unphysical_metric_expanded}
\end{equation}
The coordinate $\hat u$ is a future-directed affine parameter along the null generators of $\mathcal I^+$, and $\hat n^a = (\partial/\partial \hat{u})^a$ denotes the generator of $\mathcal I^+$. $\hat\gamma_{ab}= \hat\gamma_{ij}(dx^i)_a(dx^j)_b$ is the induced metric on the codimension-two cross sections of $\mathcal I^+$ which satisfies $\hat\gamma_{ab}\hat n^a=0$ and $\hat\gamma_{ab}(\partial/\partial\Omega)^a=0$. For a rotating background, the Killing field generating the horizon is distinguished from the asymptotically timelike Killing field $t^a$ by $\xi^a=t^a+\Omega_H\psi^a$, where $\psi^a$ is the axial Killing field and $\Omega_H$ is the horizon angular velocity. Their extensions to $\mathcal I^+$ satisfy
\begin{equation}
 \hat t^a=(\hat t^c\hat\nabla_c\hat u)\hat n^a\, , \qquad \hat\xi^a=\hat t^a+\Omega_H\hat\psi^a\, . 
\label{eq:main_xi_null_infinity_expanded}
\end{equation}
The coefficient $\hat t^c\hat\nabla_c\hat u$ is positive and constant, and the cross sections are chosen so that $\hat\psi^a\hat\nabla_a\hat u=0$. For rotating black holes, we work in an axisymmetric gauge for the background Maxwell field and restrict the perturbations on null infinity to those satisfying
\begin{equation}
 \mathcal L_\psi\delta g_{ab}=0\, , \qquad \mathcal L_\psi\delta A_a=0\, . 
\label{eq:main_axisymmetric_perturbations}
\end{equation}
Consequently, $\mathcal L_\xi\delta\phi=\mathcal L_t\delta\phi$ and $\hat\xi^a\hat\nabla_a\hat u=\hat t^a\hat\nabla_a\hat u$. Keeping $\Omega$ and the leading cross-section metric $\hat\gamma_{ab}$ fixed, we introduce the rescaled metric perturbation
\begin{equation}
 \hat\tau^{ab}=\Omega^{-1}\delta\hat g^{ab}\, , \qquad \hat\tau_{ab}=-\Omega^{-1}\delta\hat g_{ab}\, . 
\label{eq:main_rescaled_perturbation_null_infinity}
\end{equation}
Indices on $\hat\tau_{ab}$ are raised and lowered with the background $\hat g_{ab}$. The standard falloff and regularity conditions for metric perturbations on future null infinity~\cite{Hollands:2012sf,Wald:1999wa} ensure that $\hat\tau_{ab}$ extends smoothly to $\mathcal I^+$, with $\hat g^{ab}\hat\tau_{ab}=O(\Omega)$ and $\hat\tau_{ab}\hat n^a=O(\Omega)$. This convention differs by an overall sign from the rescaled covariant metric perturbation. The gravitational symplectic-current flux through $\mathcal I^+_{12}$ is determined by the Bondi news tensor~\cite{Wald:1999wa,Hollands:2012sf}, and takes the form
\begin{equation}
  \int_{\mathcal{I}^+_{12}}\hat{\boldsymbol\omega}^{GR} \overset{\mathcal I^+}{=} \frac{1}{16\pi G} \int_{\mathcal{I}^+_{12}} \hat{\boldsymbol\epsilon}^{(3)} (\hat t^c\hat\nabla_c\hat u) \delta\hat N^{ab}\delta\hat N_{ab} - \frac{1}{32\pi G} \Delta\int_{\hat{\mathcal C}} \hat{\boldsymbol\epsilon}^{(2)} (\hat t^c\hat\nabla_c\hat u) \hat\tau^{ab}\delta\hat N_{ab}\, .
\label{eq:main_gr_null_infinity_flux_expanded}
\end{equation}
Here, $\hat{\boldsymbol\epsilon}^{(3)}$ is the volume $3$-form on $\mathcal I^+$, and $\hat{\boldsymbol\epsilon}^{(2)}$ is the area $2$-form on $\hat{\mathcal C}$, which is a cross section of $\mathcal I^+$. The electromagnetic symplectic-current flux through $\mathcal I^+_{12}$ can be written as
\begin{equation}
  \int_{\mathcal{I}^+_{12}}\hat{\boldsymbol\omega}^{EM} \overset{\mathcal I^+}{=} -\frac{1}{4\pi G} \Delta\int_{\hat{\mathcal C}} \hat{\boldsymbol{\epsilon}}^{(2)} (\hat t^c\hat\nabla_c\hat u) \hat n_a\delta\hat A_b\delta\hat F^{ab} + \int_{\mathcal{I}^+_{12}} \hat{\boldsymbol{\epsilon}}^{(3)} (\hat t^c\hat\nabla_c\hat u) \hat n^a\hat n^b\delta^2\hat T^{EM}_{ab}\, .
\label{eq:main_em_null_infinity_flux_expanded}
\end{equation}
The constraint term $\hat\xi^a\delta^2\hat{\boldsymbol C}_a(\hat\phi)$ has no contribution. Since the background radiative electromagnetic field vanishes, the electromagnetic flux has no first-order contribution, $\hat n^a\hat n^b\delta\hat T^{EM}_{ab}\overset{\mathcal I^+}{=}0$. The contribution of Eq.~\eqref{eq:main_modified_current_expanded} on $\mathcal{I}^+$ is therefore 
\begin{equation}
\begin{aligned}
  \int_{\mathcal{I}^+_{12}}\hat{\boldsymbol\omega}' \overset{\mathcal I^+}{=}\;& \frac{1}{16\pi G} \int_{\mathcal{I}^+_{12}} \hat{\boldsymbol\epsilon}^{(3)} (\hat t^c\hat\nabla_c\hat u) \delta\hat N^{ab}\delta\hat N_{ab}  \\
  &- \frac{1}{32\pi G} \Delta\int_{\hat{\mathcal C}} \hat{\boldsymbol\epsilon}^{(2)} (\hat t^c\hat\nabla_c\hat u) \hat\tau^{ab}\delta\hat N_{ab} \\
  &- \frac{1}{4\pi G} \Delta\int_{\hat{\mathcal C}} \hat{\boldsymbol{\epsilon}}^{(2)} (\hat t^c\hat\nabla_c\hat u) \hat n_a\delta\hat A_b\delta\hat F^{ab}  \\
  &+ \int_{\mathcal{I}^+_{12}} (\hat t^c\hat\nabla_c\hat u) \left[ \delta\hat{\boldsymbol{\epsilon}}^{(3)} 2\hat n^a\hat n^b\delta\hat T^{EM}_{ab} + \hat{\boldsymbol{\epsilon}}^{(3)} \hat n^a\hat n^b\delta^2\hat T^{EM}_{ab} \right]\, .
\end{aligned}
\label{eq:main_omega_prime_null_infinity_flux_expanded}
\end{equation}
Following the scheme in~\cite{Wald:1999wa}, after canceling the boundary terms in the expression above, the modified canonical energy can be defined as
\begin{equation}
\begin{aligned}
  \mathcal{E}'[\Sigma(t)] \equiv\;& \mathcal{E}[\Sigma(t)] + \frac{1}{32\pi G} \int_{\mathcal{C}}\boldsymbol {\epsilon}^{(2)} (\xi^c\nabla_c v) \left( \delta \gamma^{ab}\partial_v \delta \gamma_{ab} + 2\delta \theta_v \gamma^{ab}\delta \gamma_{ab} \right) \\
  &- \frac{1}{32\pi G} \int_{\hat{\mathcal C}}\hat{\boldsymbol\epsilon}^{(2)} (\hat t^c\hat\nabla_c\hat u) \hat\tau^{ab}\delta\hat N_{ab} \\
  &- \frac{1}{4\pi G} \int_{\mathcal C} (\xi^c\nabla_c v)\boldsymbol{\epsilon}^{(2)} k_a\delta A_b\delta F^{ab} \\
  &- \frac{1}{4\pi G} \int_{\hat{\mathcal C}} (\hat t^c\hat\nabla_c\hat u) \hat{\boldsymbol{\epsilon}}^{(2)} \hat n_a\delta\hat A_b\delta\hat F^{ab} \, .
\end{aligned}
\label{eq:modified_canonical_energy_prime_definition}
\end{equation}
Consequently, the change in $\mathcal E'[\Sigma(t)]$ between $\Sigma(t_1)$ and $\Sigma(t_2)$ is
\begin{equation}
\begin{aligned}
  {\mathcal{E}'}[\Sigma(t_2)] - {\mathcal{E}'}[\Sigma(t_1)] ={}& -\frac{1}{4\pi G} \int_{\mathcal H^+_{12}} \boldsymbol {\epsilon}^{(3)} (\xi^c\nabla_c v) \left[ \delta\sigma^{ab}\delta\sigma_{ab} - \frac{1}{2}(\delta\theta_v)^2 \right] \\
  &- \int_{\mathcal{H}^+_{12}} (\xi^c\nabla_c v) \left[ \delta \boldsymbol {\epsilon}^{(3)} 2 k^a k^b\delta T'_{ab} + \boldsymbol {\epsilon}^{(3)} k^a k^b\delta^2 T'_{ab} \right] \\
  &- \frac{1}{16\pi G} \int_{\mathcal{I}^+_{12}} \hat{\boldsymbol\epsilon}^{(3)} (\hat t^c\hat\nabla_c\hat u) \delta\hat N^{ab}\delta\hat N_{ab} \\
  &- \int_{\mathcal{I}^+_{12}} (\hat t^c\hat\nabla_c\hat u) \left[ \delta\hat{\boldsymbol{\epsilon}}^{(3)} 2\hat n^a\hat n^b\delta\hat T^{EM}_{ab} + \hat{\boldsymbol{\epsilon}}^{(3)} \hat n^a\hat n^b\delta^2\hat T^{EM}_{ab} \right]\, .
\end{aligned}
\label{eq:canonical_energy_entropy_relation}
\end{equation}
The first and third terms are the gravitational contributions associated with horizon shear and Bondi news, respectively. The second term is the constraint contribution on $\mathcal H^+$, determined by the stress-energy perturbations along $k^a$, while the last term is the corresponding Einstein--Maxwell contribution on $\mathcal I^+$.

\section{Dynamical Black Hole Entropy}
\label{sec:dynamical_entropy}
In this section, we obtain the so-called ``balance law'' in terms of the modified canonical energy and derive the second-order entropy formula. We then compare the result with the area of the apparent horizon and discuss the role of the null energy condition in establishing the area law. Here, we only provide a brief derivation, and more details are given in Appendix~\ref{app:entropy}.

A relation between the modified canonical energy and the dynamical entropy is obtained next. The boundary terms absorbed in the definition of $\mathcal E'[\Sigma(t)]$ can be equivalently described by adding exact forms on $\mathcal H^+_{12}$ and $\mathcal I^+_{12}$. On the future horizon at first order, the pullback of the symplectic potential is an exact variation~\cite{Hollands:2024vbe} $\underline{\boldsymbol\Theta}(\phi,\delta\phi) \overset{\mathcal H^+}{=} \delta\boldsymbol B_{\mathcal H^+}(\phi)$, while on future null infinity one can introduce the corresponding term through~\cite{Wald:1999wa}
\begin{equation}
  \delta\hat{\boldsymbol B}_{\mathcal I^+}(\hat\phi) \overset{\mathcal I^+}{=} \hat{\boldsymbol\Theta}(\hat\phi,\delta\hat\phi) - \hat{\boldsymbol\Theta}'(\hat\phi,\delta\hat\phi)\, .
\label{eq:main_BI_definition_expanded}
\end{equation}
We define the modified canonical energy forms as
\begin{equation}
  \boldsymbol e_\Phi \overset{\mathcal H^+}{=} \boldsymbol\omega'(\phi;\delta\phi,\mathcal L_\xi\delta\phi) + d\left[ \xi\cdot\delta\boldsymbol\Theta(\phi,\delta\phi) - \xi\cdot\delta^2\boldsymbol B_{\mathcal H^+}(\phi) \right]\, ,
\label{eq:main_ephi_horizon_definition_expanded}
\end{equation}
and
\begin{equation}
  \hat{\boldsymbol e}_\Phi \overset{\mathcal I^+}{=} \hat{\boldsymbol\omega}'(\hat\phi;\delta\hat\phi, \mathcal L_{\hat\xi}\delta\hat\phi) + d\left[ \hat\xi\cdot\delta\hat{\boldsymbol\Theta}(\hat\phi,\delta\hat\phi) - \hat\xi\cdot\delta^2\hat{\boldsymbol B}_{\mathcal I^+}(\hat\phi) \right]\, .
\label{eq:main_ephi_infinity_definition_expanded}
\end{equation}
The integrals of the exact forms at the ends of Eq.~\eqref{eq:main_ephi_horizon_definition_expanded} and Eq.~\eqref{eq:main_ephi_infinity_definition_expanded} precisely cancel the boundary integral terms in Eq.~\eqref{eq:modified_canonical_energy_prime_definition}. Thus, Eq.~\eqref{eq:canonical_energy_entropy_relation} can be rewritten as
\begin{equation}
  \mathcal E'[\Sigma(t_2)]-\mathcal E'[\Sigma(t_1)] = -\int_{\mathcal H^+_{12}}\boldsymbol e_\Phi - \int_{\mathcal I^+_{12}}\hat{\boldsymbol e}_\Phi\, .
\label{eq:main_ephi_balance_expanded}
\end{equation}

The dynamical entropy $(n-2)$-form on a horizon cross section is defined by~\cite{Hollands:2024vbe}
\begin{equation}
  \boldsymbol S = \frac{2\pi}{\kappa_3} \left[ \boldsymbol Q_\xi(\phi) - \xi\cdot\boldsymbol B_{\mathcal H^+}(\phi) \right]\, ,
\label{eq:main_entropy_form_definition_expanded}
\end{equation}
where $2\kappa_3^2 = -(\nabla_{[a}\xi_{b]})(\nabla^{[a}\xi^{b]})$ and for the stationary background $\kappa_3=\kappa$. The pullback of the Noether charge to a horizon cross section is
\begin{equation}
  \boldsymbol Q_\xi(\phi) = \frac{\kappa_3}{8\pi G}\boldsymbol\epsilon^{(2)} - \frac{1}{8\pi G} \boldsymbol\epsilon^{(4)}_{ab}F^{ab}\xi^cA_c\, .
\label{eq:main_noether_charge_em_expanded}
\end{equation}
On $\mathcal H^+$, the electromagnetic part does not contribute to the entropy variation after imposing the gauge condition. In the present gauge, we choose the electromagnetic horizon correction term to vanish through second order, leaving
\begin{equation}
  \boldsymbol B_{\mathcal H^+}(\phi) = \boldsymbol B_{\mathcal H^+}(g) = \frac{1}{8\pi G}\boldsymbol\epsilon^{(3)}\theta_v\, .
  \label{eq:BH_GR_definition}
\end{equation}
Taking the second variation of Eq.~\eqref{eq:main_entropy_form_definition_expanded} and using Eqs.~\eqref{eq:main_ephi_horizon_definition_expanded}~\eqref{eq:main_noether_charge_em_expanded} and~\eqref{eq:BH_GR_definition} gives
\begin{equation}
\frac{\kappa}{2\pi}\Delta\delta^2S = \int_{\mathcal H^+_{12}}\boldsymbol e_\Phi - \int_{\mathcal H^+_{12}}2d\left[ \delta\xi\cdot\delta\boldsymbol B_{\mathcal H^+}(g) -\frac{\delta\kappa_3}{\kappa} \xi\cdot\delta\boldsymbol B_{\mathcal H^+}(g) \right]\, .
\label{eq:main_entropy_balance_expanded}
\end{equation}
This relation is commonly known as the ``balance law''. Using the same reduction as in~\cite{Fu:2026itf}, which combines the flux with the first- and second-order Raychaudhuri equations and the shear contraction identity, we obtain
\begin{equation}
  \delta^2S = \frac{1}{4G} \delta^2\int_{\mathcal C}dA_\epsilon(1-v\theta_v) + \frac{1}{2G\kappa} \int_{\mathcal C}dA_\epsilon \left[ (v\delta\kappa_3-\delta h)\delta\theta_v \right]\, .
\label{eq:main_entropy_formula_expanded}
\end{equation}

We now discuss the consequences of Eq.~\eqref{eq:main_entropy_formula_expanded}. First, we impose the null energy condition throughout the smooth two-sided family near $s=0$ and assume that the perturbation settles down to a stationary state at late times. As in~\cite{Fu:2026itf}, for the selected locally outermost apparent-horizon branch, the null energy condition implies $U(s;v,x)\geqslant0$. Since $U(0;v,x)=0$, Fermat's theorem gives $\delta U=0$. Equation~\eqref{eq:main_delta_theta_relation_expanded} then gives $\delta\theta_v=0$, and the first-order Raychaudhuri equation $\partial_v\delta\theta_v=-8\pi G\delta T'_{ab}k^ak^b$ yields
\begin{equation}
  \delta T'_{ab}k^ak^b=0\, , \qquad \delta\theta_v=0\, .
  \label{eq:energy_momentum_condition}
\end{equation}
With $k^a$ fixed and null, the null energy condition together with $T'_{ab}k^ak^b=0$ on the background and Eq.~\eqref{eq:energy_momentum_condition} implies $\delta^2T'_{ab}k^ak^b\geqslant0$. Substituting these relations into Eq.~\eqref{eq:canonical_energy_entropy_relation}, we have
\begin{equation}
  \begin{aligned}
  {\mathcal{E}'}[\Sigma(t_2)] - {\mathcal{E}'}[\Sigma(t_1)] ={}& -\frac{1}{4\pi G} \int_{\mathcal H^+_{12}} \boldsymbol {\epsilon}^{(3)} (\xi^c\nabla_c v) \delta\sigma_{ab}\delta\sigma^{ab} \\
    &- \int_{\mathcal{H}^+_{12}} \boldsymbol {\epsilon}^{(3)} (\xi^c\nabla_c v) k^a k^b\delta^2T'_{ab} \\
    &- \frac{1}{16\pi G} \int_{\mathcal{I}^+_{12}} \hat{\boldsymbol\epsilon}^{(3)} (\hat t^c\hat\nabla_c\hat u) \delta \hat N^{ab}\delta \hat N_{ab} \\
    &- \int_{\mathcal{I}^+_{12}} (\hat t^c\hat\nabla_c\hat u) \hat{\boldsymbol{\epsilon}}^{(3)} \hat n^a\hat n^b\delta^2\hat T^{EM}_{ab} \\
    &\leqslant 0 \, .
    \label{eq:reduced_canonical_energy_entropy_relation}
  \end{aligned}
\end{equation}
According to the arguments of~\cite{Hollands:2012sf,Fu:2026itf}, the gravitational parts of the first three terms on the right-hand side are nonpositive. The electromagnetic flux contained in the second term is nonnegative by the null energy condition; hence its contribution, with the overall minus sign, is nonpositive. Furthermore, the fourth term is nonpositive because the electromagnetic radiation flux is a nonnegative squared norm with respect to the transverse metric. Thus, the modified canonical energy is nonincreasing.

Consistently, integrating Eq.~\eqref{eq:main_delta_theta_relation_expanded} over a compact cross section without boundary gives
\begin{equation}
 \int_{\mathcal C}dA_\epsilon\,\delta\theta_v =\int_{\mathcal C}dA_\epsilon\,\mu\delta U=0\, . 
\label{eq:main_integrated_first_order_marginal}
\end{equation}
This agrees with $\delta U=0$ obtained from the smooth two-sided branch above. Therefore, Eqs.~\eqref{eq:main_area_expansion_expanded} and \eqref{eq:main_entropy_formula_expanded} give
\begin{equation}
  \delta^2 S = \frac{1}{4G}\delta^2 A_{\mathcal T}(v)\, .
\end{equation}
Hence, for the smooth two-sided apparent-horizon branch specified above, under the null energy condition and the asymptotic stationary condition, the dynamical entropy is proportional to the area of the apparent horizon at second order. The flux inequality in Eq.~\eqref{eq:reduced_canonical_energy_entropy_relation} further gives the second-order form of the classical second law, $\Delta\delta^2S\geqslant0$.

If the null energy condition is not imposed, we retain the same smooth two-sided family but $U(s;v,x)\geqslant0$ is no longer guaranteed. Consequently, $\delta U$ and $\delta\theta_v$ need not vanish, and the additional geometric terms in Eq.~\eqref{eq:main_area_expansion_expanded} may remain~\cite{Fu:2026itf}. Therefore, Eq.~\eqref{eq:main_entropy_formula_expanded} is not generically proportional to the area of the apparent horizon. The area law relation may still be recovered by imposing additional conditions, for example by choosing the variation of $\xi^a$ appropriately. Such a choice, however, is an extra input and is not uniquely determined by the perturbative equations. Thus, without the null energy condition, neither the apparent horizon area law nor the second law follows automatically.

The same result can also be obtained for a one-sided perturbative family under the null energy condition and the late-time stationary assumption. Assuming the matter flux is quadratic in the perturbations, the first-order Raychaudhuri equation and late-time stationarity imply $\delta\theta_v=0$. Since $U\geqslant0$, $\mu>0$, and $\delta U$ has a definite sign on each one-sided branch, the integrated first-order marginal equation then implies $\delta U=0$, after which the derivation proceeds as in the smooth two-sided case~\cite{Fu:2026itf}.

\section{Conclusion and discussion}
\label{sec:conclusion}

In this work, we investigate the entropy of dynamical black holes in Einstein--Maxwell theory to second order. Starting from a stationary electrovacuum background with a bifurcate Killing horizon, we use Gaussian null coordinates to determine the location of the apparent horizon perturbatively in the neighborhood of the event horizon. The area of the apparent horizon is expressed in terms of the expansions and the corresponding geometric quantities.

To connect this geometric construction with black-hole thermodynamics, we employ the covariant phase space formalism and introduce a modified canonical energy that incorporates contributions from external matter. After evaluating the fluxes through $\mathcal H^+$ and $\mathcal I^+$, appropriate exact forms are added on these null boundaries to cancel the boundary contributions. The resulting balance law relates the modified canonical energy to the second-order entropy formula.

Combining the entropy formula with the geometry of the apparent horizon, we show that under the null energy condition and the asymptotic stationary condition, the first-order expansion $\delta\theta_v$ vanishes and the displacement of the apparent horizon satisfies $\delta U=0$. Therefore, the extra geometric terms in the second-order area formula vanish, and the dynamical entropy is proportional to the area of the apparent horizon, $\delta^2S=(1/4G)\delta^2A_{\mathcal T}(v)$. The same assumptions also lead to the second law at second order. Without the null energy condition, the additional geometric contributions in the second-order area formula may not vanish. The area law and the second law can be recovered by imposing an appropriate choice of the variation of $\xi^a$, but it can't be uniquely fixed by the perturbative equations.

Several questions remain open. First, it would be useful to understand whether an appropriate entropy functional can be defined when the null energy condition is violated, especially in the presence of quantum matter. Second, the uniqueness of dynamical entropy beyond linear order remains to be clarified~\cite{Hollands:2024vbe}. Since the form $\boldsymbol B_{\mathcal H^+}(\phi)$ is not uniquely determined beyond first order, additional second-order terms may contribute without affecting the linearized entropy. Although we adopt $\delta^2\boldsymbol B^{EM}_{\mathcal H^+}(\phi)=0$, this choice does not remove the general second-order ambiguity in $\boldsymbol B_{\mathcal H^+}(\phi)$. Finally, extensions to asymptotically AdS spacetimes and to more general diffeomorphism covariant theories of gravity, including higher-curvature theories, would help clarify the universality of the characterization of black-hole entropy in terms of the apparent horizon.

\section*{Acknowledgement}
This work is supported in part by the National Natural Science Foundation of China with grants No. 12475063, No. 12075232 and No. 12247103. 

\appendix

\section{Residual GNC freedom}
\label{app:residual_gnc_freedom}

\subsection{Coordinate transformations preserving the GNC form}
\label{app:gnc_preserving_transformations}

We now discuss the residual coordinate freedom. After introducing the perturbation, the metric can be written as
\begin{equation}
  ds^2 = -2du\,dv -u^2\alpha(s;u,v,x)\,dv^2 -2u\beta_i(s;u,v,x)\,dv\,dx^i +\gamma_{ij}(s;u,v,x)\,dx^i\,dx^j\, .
  \label{app:perturbed_gnc_metric}
\end{equation}
Consider a coordinate transformation of the following form
\begin{equation}
  u=a\tilde u\,, \qquad v=\frac{\tilde v}{a}\,, \qquad x^i=g^i(\tilde x)\,,
  \label{app:residual_gnc_transformation}
\end{equation}
where $a$ is a positive constant. The metric keeps the Gaussian null form under this transformation~\cite{Bhattacharyya:2021jhr,Hollands:2022fkn},
\begin{equation}
  ds^2 = -2d\tilde u\,d\tilde v -\tilde u^2\tilde\alpha(s;\tilde u,\tilde v,\tilde x)\, d\tilde v^2 -2\tilde u\,\tilde\beta_i(s;\tilde u,\tilde v,\tilde x)\, d\tilde v\,d\tilde x^i +\tilde\gamma_{ij}(s;\tilde u,\tilde v,\tilde x)\, d\tilde x^id\tilde x^j\, .
  \label{app:tilded_gnc_metric}
\end{equation}
The corresponding coefficient functions transform as
\begin{equation}
  \tilde\alpha(s;\tilde u,\tilde v,\tilde x) = \alpha(s;u,v,x)\, ,
  \label{app:alpha_residual_transform}
\end{equation}
\begin{equation}
  \tilde\beta_i(s;\tilde u,\tilde v,\tilde x) = \frac{\partial g^j}{\partial\tilde x^i} \beta_j(s;u,v,x)\, ,
  \label{app:beta_residual_transform}
\end{equation}
and
\begin{equation}
  \tilde\gamma_{ij}(s;\tilde u,\tilde v,\tilde x) = \frac{\partial g^k}{\partial\tilde x^i} \frac{\partial g^l}{\partial\tilde x^j} \gamma_{kl}(s;u,v,x)\, ,
  \label{app:gamma_residual_transform}
\end{equation}

An $x$-dependent affine rescaling of the horizon generators requires corresponding corrections to the coordinate transformation away from the horizon in order to preserve the GNC form. For a general perturbed metric, the resulting near-horizon expansions on a patch with $\tilde v\ne0$ are~\cite{Bhattacharyya:2022njk,Kar:2024dqk}
\begin{equation}
  u = e^{f(\tilde x)}\tilde u \left[1+\sum_{n=1}^{\infty}(-\tilde u\tilde v)^n X_{(n)}(s;\tilde v,\tilde x)\right] ,
  \label{app:general_residual_u_transform}
\end{equation}
\begin{equation}
  v = e^{-f(\tilde x)}\tilde v \left[1+\sum_{n=1}^{\infty}(-\tilde u\tilde v)^n Y_{(n)}(s;\tilde v,\tilde x)\right] ,
  \label{app:general_residual_v_transform}
\end{equation}
and
\begin{equation}
  x^i = \tilde x^i+\sum_{n=1}^{\infty}(-\tilde u\tilde v)^n Z^i_{(n)}(s;\tilde v,\tilde x) .
  \label{app:general_residual_x_transform}
\end{equation}
Accordingly, the coefficients may have explicit $\tilde v$ dependence. The series are understood order by order near the horizon, and the resulting coordinate transformation remains regular at $\tilde v=0$.

We construct the new coordinates for each fixed $s$. Here, $f=f(\tilde x)$ is a smooth function on the horizon cross section that parametrizes the affine rescaling of the horizon generators. In the following, $\gamma_{ij}$ and $\beta_i$ on the horizon are evaluated at $(u,v,x)=(0,e^{-f}\tilde v,\tilde x)$; set $f_i:=\partial_{\tilde x^i}f$ and $f^i:=\gamma^{ij}f_j$ there. On $u=0$, the transformation gives $\partial_{\tilde v}=e^{-f}\partial_v$ and $\partial_{\tilde x^i} =\partial_i-e^{-f}\tilde v f_i\partial_v$. Requiring the transverse coordinate vector $(\partial_{\tilde u})^a$ to be null and orthogonal to the constant-$\tilde v$ horizon cross sections, with $\tilde g_{\tilde u\tilde v}=-1$, fixes its initial components:
\begin{equation}
 \left.\frac{\partial u}{\partial\tilde u}\right|_{\tilde u=0} =e^f, \qquad \left.\frac{\partial v}{\partial\tilde u}\right|_{\tilde u=0} =\frac12e^{-f}\tilde v^2f_if^i, \qquad \left.\frac{\partial x^i}{\partial\tilde u}\right|_{\tilde u=0} =-\tilde v f^i\, . 
\label{app:residual_transverse_initial_vector}
\end{equation}
Substitution of Eq.~\eqref{app:residual_transverse_initial_vector} into the metric \eqref{app:perturbed_gnc_metric} gives $\tilde g_{\tilde u\tilde u}=\tilde g_{\tilde u i}=0$ and $\tilde g_{\tilde u\tilde v}=-1$ on the horizon. The Jacobian determinant is one there, so these are valid local coordinates.

Now use $\tilde u$ as an affine parameter along the transverse null geodesics with these initial derivatives, holding $(\tilde v,\tilde x)$ fixed. The geodesic initial-value problem uniquely determines the three functions $u(\tilde u,\tilde v,\tilde x)$, $v(\tilde u,\tilde v,\tilde x)$, and $x^i(\tilde u,\tilde v,\tilde x)$ locally. At $\tilde v=0$, the initial derivatives reduce to $\partial_{\tilde u}u=e^f$ and $\partial_{\tilde u}v=\partial_{\tilde u}x^i=0$. Since $(\partial_u)^a$ generates affinely parametrized null geodesics, the exact transformation there is $(u,v,x^i)=(e^f\tilde u,0,\tilde x^i)$.

Using $\tilde u$ as an affine parameter along the transverse null geodesics, together with the commutativity of coordinate derivatives, gives
\begin{equation}
 \partial_{\tilde u}\tilde g_{\tilde u\tilde u}=0, \qquad \partial_{\tilde u}\tilde g_{\tilde u\tilde v} =\frac12\partial_{\tilde v} \tilde g_{\tilde u\tilde u}=0, \qquad \partial_{\tilde u}\tilde g_{\tilde u i} =\frac12\partial_{\tilde x^i} \tilde g_{\tilde u\tilde u}=0\, . 
\label{app:residual_gauge_propagation}
\end{equation}
Equation~\eqref{app:residual_gauge_propagation} extends $\tilde g_{\tilde u\tilde u}=\tilde g_{\tilde u i}=0$ and $\tilde g_{\tilde u\tilde v}=-1$ from the horizon to its neighborhood. The horizon transformation in Eqs.~\eqref{app:general_residual_u_transform}--%
\eqref{app:general_residual_x_transform} also gives $\tilde g_{\tilde v\tilde v}=\tilde g_{\tilde v i}=0$ on the horizon. Since $f$ is constant along each generator, $\tilde v$ remains an affine parameter there, which implies
\begin{equation}
 \left.\partial_{\tilde u}\tilde g_{\tilde v\tilde v} \right|_{\tilde u=0} =2\left.\tilde\Gamma^{\tilde v}{}_{\tilde v\tilde v} \right|_{\tilde u=0}=0\, . 
\label{app:residual_no_linear_vv_term}
\end{equation}
Consequently, smooth functions $\tilde\alpha$, $\tilde\beta_i$, and $\tilde\gamma_{ij}$ exist such that
\begin{equation}
 \tilde g_{\tilde v\tilde v}=-\tilde u^2\tilde\alpha, \qquad \tilde g_{\tilde v i}=-\tilde u\tilde\beta_i, \qquad \tilde g_{ij}=\tilde\gamma_{ij}\, . 
\label{app:residual_remaining_components}
\end{equation}
Equations~\eqref{app:residual_gauge_propagation} and \eqref{app:residual_remaining_components} account for every component of the transformed metric and reproduce the GNC line element \eqref{app:tilded_gnc_metric} exactly. The geodesic construction fixes the transformation to every order without requiring convergence of the series in Eqs.~\eqref{app:general_residual_u_transform}--%
\eqref{app:general_residual_x_transform}. At $\tilde v=0$, the coordinate map remains regular and can instead be expanded directly in $\tilde u$.

For example, expanding the null geodesics with the Eq.~\eqref{app:residual_transverse_initial_vector} gives the first corrections
\begin{equation}
 Z^i_{(1)}=f^i, \qquad Y_{(1)}=-\frac12f_if^i, \qquad X_{(1)}=-\frac12\beta_i f^i +\frac{\tilde v}{4} \partial_{\tilde v}\!\left[ \gamma_{ij}(s;0,e^{-f}\tilde v,\tilde x) \right]f^if^j\, . 
\label{app:residual_first_coefficients}
\end{equation}
These coefficients agree with Eq.~(2.13) of~\cite{Kar:2024dqk} after identifying $r=-u$, $\rho=-\tilde u$, $\tau=\tilde v$, and $\zeta=-f$. In particular, $Z^i_{(1)}=\gamma^{ij}f_j$ generally varies with $\tilde v$ in a dynamical geometry.

When the geometry is stationary, the boost symmetry $(u,v)\mapsto(cu,v/c)$ preserves the horizon geometry. By uniqueness, the constructed coordinates respect $(\tilde u,\tilde v)\mapsto(c\tilde u,\tilde v/c)$, so $u/\tilde u$, $v/\tilde v$, and $x^i$ are functions only of $\tilde u\tilde v$ and $\tilde x$. The original product-dependent expressions are therefore recovered in that case. For a general family $g(s)$, the coordinate map may depend on $s$; variations must then be taken at fixed tilded coordinates after pulling back the entire one-parameter family. Comparing Eq.~\eqref{eq:main_area_expansion_expanded} also requires matching the same geometric cross sections, since constant-$v$ and constant-$\tilde v$ cross sections generally differ.

Hence the three transformations \eqref{app:general_residual_u_transform}--%
\eqref{app:general_residual_x_transform}, with their coefficients fixed by the null-geodesic construction, take the original metric \eqref{app:perturbed_gnc_metric} into precisely the GNC metric \eqref{app:tilded_gnc_metric}. This statement holds exactly in a neighborhood of the horizon for each fixed $s$, including the smooth extension through $\tilde v=0$.

\subsection{General residual transformations of the second-order area formula}
\label{app:area_formula_covariance}

We now discuss Eq.~\eqref{eq:main_area_expansion_expanded} under the three general residual transformations \eqref{app:general_residual_u_transform}--%
\eqref{app:general_residual_x_transform}. The metric and the marginal cross sections are transformed together, including any $s$ dependence of the coordinate map. Consider an original marginal cross section $v=v_0$, $u=U(s;v_0,x)$. Its position in the new coordinates is $\tilde u=\tilde U(s;\tilde x)$, $\tilde v=\tilde V(s;\tilde x)$, with $\tilde V(0;\tilde x)=v_0e^{f(\tilde x)}$. The coordinate transformations and Eq.~\eqref{app:residual_transverse_initial_vector} give the first-order variations
\begin{equation}
 \delta\tilde U=e^{-f}\delta U, \quad\delta\tilde V=-\frac12v_0^2e^{2f}f_if^i\delta\tilde U, \quad\delta x^i=-v_0e^f f^i\delta\tilde U\, . 
\label{app:area_transformed_cut}
\end{equation}
Let $\tilde q_{ij}(s)$ denote the components of the induced metric on this cross section in the $\tilde x^i$ coordinates. Substituting $d\tilde u=\partial_i\tilde U\,d\tilde x^i$ and $d\tilde v=\partial_i\tilde V\,d\tilde x^i$ into Eq.~\eqref{app:tilded_gnc_metric} gives
\begin{equation}
 \tilde q_{ij}(s)=\tilde\gamma_{ij} -\partial_i\tilde U\,\partial_j\tilde V -\partial_j\tilde U\,\partial_i\tilde V -\tilde U\left(\tilde\beta_i\partial_j\tilde V +\tilde\beta_j\partial_i\tilde V\right) -\tilde U^2\tilde\alpha\, \partial_i\tilde V\,\partial_j\tilde V\, . 
\label{app:area_transported_induced_metric}
\end{equation}
In the original coordinates, the same cross section has induced metric $q_{ij}(s)=\gamma_{ij}(s;U,v_0,x)$. The tensor transformation law therefore implies
\begin{equation}
 \tilde q_{ij}(s) =\frac{\partial x^k}{\partial\tilde x^i} \frac{\partial x^l}{\partial\tilde x^j}q_{kl}(s), \qquad \sqrt{\det\tilde q(s)} =\left|\det\frac{\partial x}{\partial\tilde x}\right| \sqrt{\det q(s)}\, . 
\label{app:area_induced_metric_transform}
\end{equation}
The determinant factor is precisely the Jacobian for changing integration variables, so
\begin{equation}
 \int d^{n-2}\tilde x\,\sqrt{\det\tilde q(s)} =\int d^{n-2}x\,\sqrt{\det\gamma(s;U,v_0,x)} =A_{\mathcal T}(s;v_0)\, . 
\label{app:area_coordinate_invariance}
\end{equation}
Accounting for both the metric perturbations and the changes in $\tilde U$ and $\tilde V$, we obtain
\begin{equation}
 \delta^2 A_{\mathcal T}(v_0) =\int_{\tilde{\mathcal C}_{v_0}}d\tilde A_\epsilon \Biggl[ \frac12\tilde q^{ij}\delta^2\tilde q_{ij} +\frac14\left(\tilde q^{ij}\delta\tilde q_{ij}\right)^2 -\frac12\tilde q^{ik}\tilde q^{jl} \delta\tilde q_{ij}\delta\tilde q_{kl} \Biggr]\, . 
\label{app:area_transported_second_variation}
\end{equation}
Eq.~\eqref{app:area_coordinate_invariance} gives identical areas in the two coordinate systems for the same cross section. Their expansions therefore have the same second-order coefficient $\delta^2 A_{\mathcal T}(v_0)$, even though the induced metric components change.

To obtain the analogue of Eq.~\eqref{eq:main_area_expansion_expanded} on constant-$\tilde v$ marginal cross sections, $\tilde U(s;\tilde v,\tilde x)$ must be determined from the marginality condition in the new GNC. These cross sections generally differ from the images of the original constant-$v$ cross sections. Their induced metric is $\tilde\gamma_{ij}(s;\tilde U,\tilde v,\tilde x)$. With $\tilde\lambda=\tilde\alpha+ \tilde\gamma^{ij}\tilde\beta_i\tilde\beta_j$ and $\tilde D_i$ the intrinsic covariant derivative on $\tilde{\mathcal C}$, the second variation of the area is
\begin{equation}
 \delta^2\tilde A_{\mathcal T} =\delta^2\int_{\tilde{\mathcal C}}d\tilde A_\epsilon +\int_{\tilde{\mathcal C}}d\tilde A_\epsilon\left[ \tilde\theta_u\delta^2\tilde U +2\delta\tilde U\,\delta\tilde\theta_u +\tilde\theta_u\delta\tilde U\, \tilde\gamma^{ij}\delta\tilde\gamma_{ij} +(\delta\tilde U)^2 (\partial_{\tilde u}\tilde\theta_u+\tilde\theta_u^2) \right]\, . 
\label{app:area_fixed_cut_expansion}
\end{equation}
Since the transformed metric retains the GNC form \eqref{app:tilded_gnc_metric}, the outgoing null normal to $\tilde u=\tilde U(s;\tilde v,\tilde x)$ at fixed $\tilde v$ is constructed in the same way as in the original coordinates. Expanding the corresponding marginality condition to first and second order therefore gives equations of the same form as Eqs.~\eqref{eq:main_delta_theta_relation_expanded} and \eqref{eq:main_delta2_theta_relation_expanded}, with all quantities defined in the tilded coordinates. Using these equations together with the background identities $\tilde\theta_v=0$ and $\tilde\theta_u=\tilde v\partial_{\tilde u}\tilde\theta_v$, and integrating by parts with no boundary contribution, we obtain
\begin{equation}
\begin{aligned}
  &\int_{\tilde{\mathcal C}}d\tilde A_\epsilon \left[\tilde\theta_u\delta^2\tilde U +\tilde\theta_u\delta\tilde U\, \tilde\gamma^{ij}\delta\tilde\gamma_{ij}\right] =-\delta^2\int_{\tilde{\mathcal C}} d\tilde A_\epsilon\,\tilde v\tilde\theta_v\\
  &\quad+\int_{\tilde{\mathcal C}}d\tilde A_\epsilon\Bigl[ \tilde\lambda\tilde v\tilde\theta_u(\delta\tilde U)^2 -\tilde v(\tilde D\delta\tilde U)^2\tilde\theta_u -2\tilde v\delta\tilde U\, \partial_{\tilde u}\delta\tilde\theta_v -\tilde v(\delta\tilde U)^2 \partial_{\tilde u}^2\tilde\theta_v -2\tilde v\delta\tilde U (\tilde D^i\delta\tilde U +\tilde\beta^i\delta\tilde U) \tilde D_i\tilde\theta_u \Bigr]\, .
\end{aligned}
\label{app:area_marginality_elimination}
\end{equation}
Substituting Eq.~\eqref{app:area_marginality_elimination} into Eq.~\eqref{app:area_fixed_cut_expansion}, eliminates the terms involving $\delta^2\tilde U$ and $\delta\tilde\gamma_{ij}$. We obtain
\begin{equation}
\begin{aligned}
  \delta^2\tilde A_{\mathcal T}(\tilde v) ={}&\delta^2\int_{\tilde{\mathcal C}} d\tilde A_\epsilon(1-\tilde v\tilde\theta_v) +\int_{\tilde{\mathcal C}}d\tilde A_\epsilon\Biggl\{ \left(1+\frac{\tilde\lambda\tilde v}{\tilde\theta_u}\right) (\tilde\theta_u\delta\tilde U)^2 -\tilde v(\tilde D\delta\tilde U)^2\tilde\theta_u\\
     &+2\delta\tilde U\left( \delta\tilde\theta_u -\tilde v\partial_{\tilde u}\delta\tilde\theta_v \right) +(\delta\tilde U)^2\left( \partial_{\tilde u}\tilde\theta_u -\tilde v\partial_{\tilde u}^2\tilde\theta_v \right)-2\tilde v\delta\tilde U\left( \tilde D^i\delta\tilde U+\tilde\beta^i\delta\tilde U \right)\tilde D_i\tilde\theta_u \Biggr\}\, .
\end{aligned}
\label{app:tilded_area_formula}
\end{equation}
It has exactly the form of Eq.~\eqref{eq:main_area_expansion_expanded} in the new GNC. Since constant-$v$ and constant-$\tilde v$ marginal cross sections generally differ, their area variations need not be equal. Equation~\eqref{app:area_coordinate_invariance} guarantees equal area variations when the same cross sections are described in the two coordinate systems.

\section{Derivation of the modified-canonical energy fluxes}
\label{app:modified_canonical_energy}

\subsection{Symplectic current form}

For the gravitational part on the horizon, we do not repeat the detailed calculation. Using the result derived in~\cite{Fu:2026itf}, one has
\begin{equation}
  \boldsymbol \omega^{GR}(g;\delta g,\mathcal{L}_\xi \delta g) \overset{\mathcal{H}^+}{=} \frac{\kappa}{32\pi G}\boldsymbol{\epsilon}^{(3)} \left\{ (\delta \gamma^{ab} +\gamma^{ab}\gamma^{cd}\delta \gamma_{cd}) \partial_v(v\partial_v \delta \gamma_{ab}) -v\left[ \partial_v\delta\gamma^{ab}\partial_v\delta\gamma_{ab} +4(\delta\theta_v)^2 \right] \right\}\, .
  \label{app:omega_GR_on_horizon}
\end{equation}
For the electromagnetic contribution, Eq.~(102) of~\cite{Sorce:2017dst} gives the corresponding symplectic current. After pulling it back to $\mathcal H^+$, we obtain
\begin{equation}
\begin{aligned}
\boldsymbol{\omega}^{EM}(\phi;\delta\phi,\mathcal{L}_\xi \delta\phi) \overset{\mathcal{H}^+}{=}& \frac{1}{4\pi G}\boldsymbol{\epsilon}^{(4)}_a \left( \delta A_b\mathcal{L}_{\xi}\delta F^{ab} -\delta F^{ab}\mathcal{L}_{\xi}\delta A_b \right) + \frac{1}{4\pi G}F^{ab} \left[ \delta A_b\mathcal{L}_{\xi}\delta\boldsymbol{\epsilon}^{(4)}_a -\delta\boldsymbol{\epsilon}^{(4)}_a\mathcal{L}_{\xi}\delta A_b \right]  \\
=& \frac{1}{4\pi G}\boldsymbol{\epsilon}^{(4)}_a \left( \delta A_b\mathcal{L}_{\xi}\delta F^{ab} -\delta F^{ab}\mathcal{L}_{\xi}\delta A_b \right) + \frac{1}{8\pi G}\boldsymbol{\epsilon}^{(3)}k_aF^{ab}\gamma^{cd} \left[ \delta\gamma_{cd}\mathcal{L}_{\xi}\delta A_b -\delta A_b\mathcal{L}_{\xi}\delta\gamma_{cd} \right]\, .
\end{aligned}
\label{app:em_current_general}
\end{equation}
Here, the horizon volume-form decomposition and the first-order variation of the induced metric have been used. To simplify the second term in Eq.~\eqref{app:em_current_general}, we use the stationarity of the background on $\mathcal{H}^+$. Since the expansion and shear of the null generators vanish there, taking the Raychaudhuri equation into account gives
\begin{equation}
  T^{EM}_{ab}k^a k^b = \frac{1}{4\pi G}\left(F_a{}^c F_{bc} - \frac{1}{4}g_{ab}F_{cd}F^{cd}\right)k^ak^b \overset{\mathcal H^+}{=} 0 \, .
  \label{app:energy_momentum_tensor}
\end{equation}
Equivalently, $F_{ac}F^{c}{}_{b}k^ak^b\overset{\mathcal H^+}{=}0$. Thus the vector $F_{ab}k^a$ has vanishing norm. Since it is also orthogonal to $k^b$ by the antisymmetry of $F_{ab}$, it must be proportional to $k_b$ on $\mathcal H^+$. Using the above result, one obtains $\delta T^{EM}_{ab}k^a k^b \overset{\mathcal H^+}{=}0$. We then impose the horizon gauge condition~\cite{Sorce:2017dst}
\begin{equation}
   k^a\delta^m A_a\overset{\mathcal H^+}{=}0 \, , \qquad (m = 0,1,2) \, .
   \label{app:horizon_EM_gauge_condition}
\end{equation}
This condition can be imposed order by order by a Maxwell gauge transformation $\delta^m A_a\to \delta^m A_a-\nabla_a\chi^{(m)}$, where $\chi^{(m)}$ denotes the $m$-th order gauge parameter with $k^a\nabla_a\chi^{(m)}\overset{\mathcal H^+}{=}k^a\delta^m A_a$.
\begin{equation}
  \mathcal L_\xi\left(k^a\delta A_a\right) =\left(\mathcal L_\xi k^a\right)\delta A_a+k^a\mathcal L_\xi\delta A_a \overset{\mathcal H^+}{=}0 \, .
  \label{app:gauge_condition_lie_derivative}
\end{equation}
Using $\mathcal L_\xi k^a\overset{\mathcal H^+}{=}-\kappa k^a$, we further obtain
\begin{equation}
  k^a\mathcal L_\xi\delta A_a \overset{\mathcal H^+}{=}0 \, .
  \label{app:gauge_condition_result}
\end{equation}
According to Eqs.~\eqref{app:horizon_EM_gauge_condition},~\eqref{app:gauge_condition_lie_derivative}, and \eqref{app:gauge_condition_result}, it follows that the second term in Eq.~\eqref{app:em_current_general} vanishes on $\mathcal H^+$. The first term can be rearranged using Cartan's formula as
\begin{equation}
  \boldsymbol{\omega}^{EM}(\phi;\delta\phi,\mathcal{L}_\xi \delta\phi) \overset{\mathcal H^+}{=} -\frac{1}{4\pi G} d\left[ (\xi^c\nabla_cv)\boldsymbol\epsilon^{(2)} k_a\delta A_b\delta F^{ab} \right] + \frac{1}{2\pi G} (\xi^c\nabla_cv)\boldsymbol\epsilon^{(3)} k^ak^b\delta F_a{}^c\delta F_{bc}\, .
\label{app:em_horizon_boundary_flux}
\end{equation}
If the integration region is chosen with the initial cross section at $v_1=0$ and the final cross section taken to the asymptotic limit $v_2\to\infty$, then the first term gives no contribution. This is because the initial cross section lies on the bifurcation surface, where the background is stationary, and we assume that the spacetime also settles down to a stationary state as $v\to\infty$. Thus both boundary cross sections are non-radiative stationary regions, so the boundary contribution vanishes. In this case, the total contribution is entirely given by the radiative flux term, which can be interpreted as the Poynting flux. For arbitrary finite values of $v_1$ and $v_2$, however, the boundary term is generally nonzero.

The electromagnetic stress tensor is
\begin{equation}
  T^{EM}_{ab} = \frac{1}{4\pi G} \left( F_{ac}F_b{}^c-\frac14g_{ab}F_{cd}F^{cd} \right)\, .
\label{app:em_stress_tensor}
\end{equation}
Since $k^a$ is fixed by~\eqref{eq:main_kl_fixed_expanded}, substituting Eq.~\eqref{app:em_stress_tensor} into Eq.~\eqref{app:em_horizon_boundary_flux}, one obtains
\begin{equation}
\boldsymbol{\omega}^{EM}(\phi;\delta\phi,\mathcal{L}_\xi \delta\phi) \overset{\mathcal H^+}{=} -\frac{1}{4\pi G} d\left[ (\xi^c\nabla_cv)\boldsymbol\epsilon^{(2)} k_a\delta A_b\delta F^{ab} \right] + (\xi^c\nabla_cv)\boldsymbol\epsilon^{(3)} k^ak^b\delta^2T^{EM}_{ab}\, .
\label{app:em_horizon_boundary_flux_stress_tensor}
\end{equation}
Then, the second and third terms in Eq.~\eqref{eq:main_modified_current_expanded} vanish after being pulled back to $\mathcal H^+$. Therefore, we only need to evaluate the constraint contribution. Using Eq.~\eqref{eq:full_Ca_expression} and assuming that the electromagnetic field satisfies the Maxwell equation and its linearized equation, the constraint contribution becomes
\begin{equation}
  \xi^a\delta^2\boldsymbol C_a(\phi) \overset{\mathcal H^+}{=} -(\xi^c\nabla_c v) \left[ \delta\boldsymbol\epsilon^{(3)} 2k^ak^b\delta T_{ab} + \boldsymbol\epsilon^{(3)} k^ak^b\delta^2T_{ab} \right] -\frac{1}{4\pi G} (\xi^a A_a)\boldsymbol\epsilon^{(4)}_c \delta^2(\nabla_bF^{bc})\, .
\label{app:main_second_order_Ca_matter_contribution}
\end{equation}
The last term in Eq.~\eqref{app:main_second_order_Ca_matter_contribution} gives no contribution on $\mathcal H^+$, since its pullback vanishes upon using $\boldsymbol\epsilon^{(4)}_a\overset{\mathcal H^+}{=} -k_a\boldsymbol\epsilon^{(3)}$ together with Eq.~\eqref{app:horizon_EM_gauge_condition}. Therefore,
\begin{equation}
  \xi^a\delta^2\boldsymbol C_a(\phi) \overset{\mathcal H^+}{=} -(\xi^c\nabla_c v) \left[ \delta\boldsymbol\epsilon^{(3)} 2k^ak^b\delta T_{ab} + \boldsymbol\epsilon^{(3)} k^ak^b\delta^2T_{ab} \right]\, .
\label{app:main_second_order_Ca_gravity_contribution}
\end{equation}

Near $\mathcal I^+$, we work with the unphysical fields
\begin{equation}
  \hat g_{ab}=\Omega^2 g_{ab}\, , \qquad \hat A_a=A_a\, , \qquad \hat F_{ab}=F_{ab}\, .
  \label{app:unphysical_GR_EM_fields}
\end{equation}
The gravitational contribution has already been evaluated in~\cite{Fu:2026itf} using the symplectic structure on $\mathcal I^+$. In the convention of Eq.~\eqref{eq:main_rescaled_perturbation_null_infinity}, the gravitational symplectic potential on $\mathcal I^+$ and the linearized Bondi news satisfy
\begin{equation}
 \hat{\boldsymbol\Theta}'{}^{GR}(\hat g,\delta\hat g) \overset{\mathcal I^+}{=} \frac{1}{32\pi G}\hat{\boldsymbol\epsilon}^{(3)} \hat N^{ab}\hat\tau_{ab}\, , \qquad\delta\hat N_{ab} \overset{\mathcal I^+}{=} \hat\gamma_a{}^c\hat\gamma_b{}^d \mathcal L_{\hat n}\hat\tau_{cd}\, . 
\label{app:gr_boundary_potential_and_news}
\end{equation}
By definition of the symplectic potential,
\begin{equation}
\begin{aligned}
\hat{\boldsymbol\omega}^{GR}(\hat g;\delta_1\hat g,\delta_2\hat g) \overset{\mathcal I^+}{=} \delta_1\hat{\boldsymbol\Theta}'{}^{GR}(\hat g,\delta_2\hat g) -\delta_2\hat{\boldsymbol\Theta}'{}^{GR}(\hat g,\delta_1\hat g)\, ,
\end{aligned}
\label{app:gr_boundary_symplectic_current}
\end{equation}
which, for vanishing background news and with Eq.~\eqref{eq:main_axisymmetric_perturbations} imposed in rotating backgrounds, reduces to
\begin{equation}
  \hat{\boldsymbol\omega}^{GR} (\hat g;\delta\hat g,\mathcal L_{\hat\xi}\delta\hat g) \overset{\mathcal I^+}{=} \frac{1}{16\pi G} \hat{\boldsymbol\epsilon}^{(3)} (\hat t^c\hat\nabla_c\hat u) \delta\hat N^{ab}\delta\hat N_{ab} - \frac{1}{32\pi G} d\left[ \hat{\boldsymbol\epsilon}^{(2)} (\hat t^c\hat\nabla_c\hat u) \hat\tau^{ab}\delta\hat N_{ab} \right]\, .
\label{app:gr_infinity_current}
\end{equation}
After integration over $\mathcal I^+_{12}$, this gives Eq.~\eqref{eq:main_gr_null_infinity_flux_expanded}. For the electromagnetic part, we first rewrite the physical current in terms of the unphysical fields defined in Eq.~\eqref{app:unphysical_GR_EM_fields}. In four dimensions, keeping $\delta\Omega=0$, Eq.~\eqref{app:em_current_general} becomes
\begin{equation}
  \hat{\boldsymbol{\omega}}^{EM} (\hat\phi;\delta\hat\phi,\mathcal L_{\hat\xi}\delta\hat\phi) = \frac{1}{4\pi G} \hat{\boldsymbol\epsilon}^{(4)}_a \left[ \delta\hat A_b \mathcal L_{\hat\xi}\delta\hat F^{ab} - \delta\hat F^{ab} \mathcal L_{\hat\xi}\delta\hat A_b \right] + \frac{1}{4\pi G} \hat F^{ab} \left[ \delta\hat A_b \mathcal L_{\hat\xi}\delta\hat{\boldsymbol\epsilon}^{(4)}_a - \delta\hat{\boldsymbol\epsilon}^{(4)}_a \mathcal L_{\hat\xi}\delta\hat A_b \right]\, .
\label{app:em_current_conformal}
\end{equation}
Here all indices are raised with $\hat g^{ab}$. Since $\delta\hat g_{ab}=O(\Omega)$ for asymptotically flat perturbations~\cite{Hollands:2012sf}, one has $\delta\hat{\boldsymbol\epsilon}^{(4)}_a=O(\Omega)$. Therefore, the second term of Eq.~\eqref{app:em_current_conformal} vanishes in the limit to $\mathcal I^+$~\cite{Bonga:2019bim}.

On $\mathcal I^+$, we impose the gauge conditions:
\begin{equation}
  \hat n^a\delta^m\hat A_a\overset{\mathcal I^+}{=}0 \, , \qquad (m = 1,2) \, .
\label{app:infinity_maxwell_gauge}
\end{equation}
The horizon gauge~\eqref{app:horizon_EM_gauge_condition} sets $k^aA_a=0$ on $\mathcal H^+$, while a charged black hole can still have $\hat n^a\hat A_a\neq0$ on $\mathcal I^+$. The $m=1,2$ conditions can be imposed by gauge transformations smooth between the two boundaries. Rewriting the electromagnetic symplectic current as the sum of a boundary term and a flux term and using Eq.~\eqref{eq:main_xi_null_infinity_expanded}, we obtain
\begin{equation}
  \hat{\boldsymbol{\omega}}^{EM} (\hat\phi;\delta\hat\phi,\mathcal L_{\hat\xi}\delta\hat\phi) \overset{\mathcal I^+}{=} -\frac{1}{4\pi G} d\left[ (\hat t^c\hat\nabla_c\hat u) \hat{\boldsymbol\epsilon}^{(2)} \hat n_a\delta\hat A_b\delta\hat F^{ab} \right] + (\hat t^c\hat\nabla_c\hat u) \hat{\boldsymbol\epsilon}^{(3)} \hat n^a\hat n^b\delta^2\hat T^{EM}_{ab}\, .
\label{app:em_infinity_current}
\end{equation}
The argument below Eq.~\eqref{app:em_horizon_boundary_flux} applies in the same way. Hence, under similar stationary boundary conditions, the boundary term gives no contribution and the flux contribution on $\mathcal I^+$ is entirely given by the Poynting flux. For arbitrary finite boundary cross sections, the boundary term must be kept explicitly.

Since the stationary background has no electromagnetic radiation through $\mathcal I^+$, an argument analogous to the discussion below Eq.~\eqref{app:energy_momentum_tensor} yields $\delta\hat T^{EM}_{ab}\hat n^a\hat n^b\overset{\mathcal I^+}{=}0$. Combining Eqs.~\eqref{app:gr_infinity_current} and \eqref{app:em_infinity_current} gives Eq.~\eqref{eq:main_omega_prime_null_infinity_flux_expanded}. Similarly, the second and third terms in Eq.~\eqref{eq:main_modified_current_expanded} vanish after being pulled back to $\mathcal I^+$. It remains to examine the constraint contribution, which takes the form
\begin{equation}
  \hat{\xi}^a \delta^2\hat{\boldsymbol C}_a(\hat\phi) = -(\hat t^b\hat\nabla_b\hat u) \left[ \delta \hat{\boldsymbol {\epsilon}}^{(3)} 2\hat n^a\hat n^b\delta \hat T_{ab} + \hat{\boldsymbol {\epsilon}}^{(3)} \hat n^a\hat n^b\delta^2\hat T_{ab} \right] -\frac{1}{4\pi G} (\hat\xi^a\hat A_a)\hat{\boldsymbol\epsilon}^{(4)}_c \delta^2(\hat\nabla_b\hat F^{bc})\, .
\label{eq:second_order_Ca_matter_contribution_Iplus}
\end{equation}
For the perturbations considered here, the external matter is assumed to cross $\mathcal H^+$ and decay sufficiently rapidly toward $\mathcal I^+$, so that $\delta\hat T_{ab}\overset{\mathcal I^+}{=}0$ and $\delta^2\hat T_{ab}\overset{\mathcal I^+}{=}0$. Consistent with the electrically neutral external matter considered here, we impose no electric-charge flux through $\mathcal I^+$ at second order. This allows electromagnetic radiation, which carries energy but no electric charge:
\begin{equation}
\hat n_c\,\delta^2(\hat\nabla_b\hat F^{bc}) \overset{\mathcal I^+}{=}0\, .
\label{app:no_second_order_charge_flux}
\end{equation}
Using $\hat{\boldsymbol\epsilon}^{(4)}_c\overset{\mathcal I^+}{=}-\hat n_c\hat{\boldsymbol\epsilon}^{(3)}$, it eliminates the last term in Eq.~\eqref{eq:second_order_Ca_matter_contribution_Iplus}. Hence the constraint term does not contribute on $\mathcal I^+$:
\begin{equation}
    \hat{\xi}^a \delta^2\hat{\boldsymbol C}_a(\hat\phi) \overset{\mathcal I^+}{=}0\, .
\end{equation}

\subsection{Absorption of boundary terms}

We now show how the boundary contributions in Eq.~\eqref{eq:modified_canonical_energy_prime_definition} are canceled by adding exact forms on the corresponding null boundaries. On $\mathcal H^+$ and $\mathcal I^+$, the relevant exact forms are
\begin{equation}
  d\left[\xi\cdot\delta\boldsymbol\Theta(\phi,\delta\phi)-\xi\cdot\delta^2\boldsymbol B_{\mathcal H^+}(\phi)\right]\, ,
\label{app:exactform_horizon}
\end{equation}
and
\begin{equation}
  d\left[\hat\xi\cdot\delta\hat{\boldsymbol\Theta}(\hat\phi,\delta\hat\phi)-\hat\xi\cdot\delta^2\hat{\boldsymbol B}_{\mathcal I^+}(\hat\phi)\right]\, .
\label{app:exactform_infinity}
\end{equation}
On the stationary background, $\boldsymbol B_{\mathcal H^+}(\phi)$ satisfies the first-order relation 
\begin{equation}
    \underline{\boldsymbol\Theta}(\phi,\delta\phi)\overset{\mathcal H^+}{=}\delta\boldsymbol B_{\mathcal H^+}(\phi)\, , 
    \label{app:horizon_counterterm_definition}
\end{equation}
where the underline denotes pullback to $\mathcal H^+$. This relation fixes $\boldsymbol B_{\mathcal H^+}(\phi)$ only to first order; its extension to second order must be specified separately~\cite{Hollands:2024vbe}. On $\mathcal I^+$, the correction term is defined through
\begin{equation}
  \delta\hat{\boldsymbol B}_{\mathcal I^+}(\hat\phi)\overset{\mathcal I^+}{=}\hat{\boldsymbol\Theta}(\hat\phi,\delta\hat\phi)-\hat{\boldsymbol\Theta}'(\hat\phi,\delta\hat\phi)\, ,
\label{app:null_infinity_counterterm_definition}
\end{equation}
where $\hat{\boldsymbol\Theta}'(\hat\phi,\delta\hat\phi)$ is a symplectic potential for the pullback of the symplectic current $\hat{\boldsymbol\omega}(\hat\phi;\delta_1\hat\phi,\delta_2\hat\phi)$ to $\mathcal I^+$.

We first evaluate Eq.~\eqref{app:exactform_horizon}. It splits into gravitational and electromagnetic parts:
\begin{equation}
d\left[\xi\cdot\delta\boldsymbol\Theta(\phi,\delta\phi)-\xi\cdot\delta^2\boldsymbol B_{\mathcal H^+}(\phi)\right]=d\left[\xi\cdot\delta\boldsymbol\Theta^{GR}(g,\delta g)-\xi\cdot\delta^2\boldsymbol B^{GR}_{\mathcal H^+}(g)\right]+d\left[\xi\cdot\delta\boldsymbol\Theta^{EM}(\phi,\delta\phi)-\xi\cdot\delta^2\boldsymbol B^{EM}_{\mathcal H^+}(\phi)\right]\, .
\label{app:exactform_horizon_split}
\end{equation}
For the gravitational part, the corresponding calculation has already been given in~\cite{Fu:2026itf}. In the present notation, it gives
\begin{equation}
\begin{aligned}
    d\left[ \xi\cdot\delta\boldsymbol\Theta^{GR}(g,\delta g) - \xi\cdot\delta^2\boldsymbol B^{GR}_{\mathcal H^+}(g) \right] \overset{\mathcal H^+}{=} &-\frac{\kappa}{32\pi G} \boldsymbol\epsilon^{(3)} \left( 2\gamma^{ab}\delta\gamma_{ab}\delta\theta_v + \delta\gamma^{ab}\partial_v\delta\gamma_{ab} \right) \\
    &-\frac{\kappa v}{32\pi G} \boldsymbol\epsilon^{(3)} \partial_v \left( 2\gamma^{ab}\delta\gamma_{ab}\delta\theta_v + 2\delta^2\gamma^{ab}\partial_v\gamma_{ab} + \delta\gamma^{ab}\partial_v\delta\gamma_{ab} \right)\, .
\end{aligned}
\label{app:boundary_term_on_horizon_GR}
\end{equation}
For the electromagnetic part, the symplectic potential is
\begin{equation}
  \boldsymbol\Theta^{EM}(\phi,\delta\phi)=-\frac{1}{4\pi G}\boldsymbol\epsilon^{(4)}_aF^{ab}\delta A_b\, .
\label{app:theta_EM_horizon}
\end{equation}
On the stationary background, $k_aF^{ab}$ is proportional to $k^b$, as shown below Eq.~\eqref{app:energy_momentum_tensor}. Together with $\boldsymbol\epsilon^{(4)}_a\overset{\mathcal H^+}{=}-k_a\boldsymbol\epsilon^{(3)}$ and the horizon gauge \eqref{app:horizon_EM_gauge_condition}, this gives $\underline{\boldsymbol\Theta}^{EM}(\phi,\delta\phi)\overset{\mathcal H^+}{=}0$. Since the minimally coupled Maxwell field gives no independent contribution to the stationary Wald entropy, we choose $\boldsymbol B^{EM}_{\mathcal H^+}(\phi)=0$ through second order. Hence $\delta^2\boldsymbol B^{EM}_{\mathcal H^+}(\phi)=0$, while $\delta\underline{\boldsymbol\Theta}^{EM}(\phi,\delta\phi)$ need not vanish and gives the boundary term below. The electromagnetic exact form then reduces to
\begin{equation}
  d\left[\xi\cdot\delta\boldsymbol\Theta^{EM}(\phi,\delta\phi)-\xi\cdot\delta^2\boldsymbol B^{EM}_{\mathcal H^+}(\phi)\right]\overset{\mathcal H^+}{=}\frac{1}{4\pi G}d\left[\boldsymbol\epsilon^{(2)}\kappa v k_a\delta A_b\delta F^{ab}\right]\, .
\label{app:exactform_horizon_EM}
\end{equation}

Integrating Eq.~\eqref{app:exactform_horizon_split} over $\mathcal H^+_{12}$ gives
\begin{equation}
\begin{aligned}
    \int_{\mathcal H^+_{12}}d\left[\xi\cdot\delta\boldsymbol\Theta(\phi,\delta\phi)-\xi\cdot\delta^2\boldsymbol B_{\mathcal H^+}(\phi)\right]=-&\frac{1}{32\pi G}\Delta\int_{\mathcal C}\boldsymbol\epsilon^{(2)}(\xi^c\nabla_c v)\left(\delta\gamma^{ab}\partial_v\delta\gamma_{ab}+2\delta\theta_v\gamma^{ab}\delta\gamma_{ab}\right)\\+&\frac{1}{4\pi G}\Delta\int_{\mathcal C}\boldsymbol\epsilon^{(2)}(\xi^c\nabla_c v)k_a\delta A_b\delta F^{ab}\, .
\end{aligned}
\label{app:horizon_exact_form_cancellation}
\end{equation}
The first term cancels the gravitational boundary contribution in Eq.~\eqref{app:omega_GR_on_horizon}, while the second cancels the electromagnetic boundary contribution in Eq.~\eqref{app:em_horizon_boundary_flux}. The same analysis is then carried out on $\mathcal I^+$. In this case, Eq.~\eqref{app:exactform_infinity} decomposes as
\begin{equation}
  d\left[\hat\xi\cdot\delta\hat{\boldsymbol\Theta}(\hat\phi,\delta\hat\phi)-\hat\xi\cdot\delta^2\hat{\boldsymbol B}_{\mathcal I^+}(\hat\phi)\right]=d\left[\hat\xi\cdot\delta\hat{\boldsymbol\Theta}^{GR}(\hat g,\delta\hat g)-\hat\xi\cdot\delta^2\hat{\boldsymbol B}^{GR}_{\mathcal I^+}(\hat g)\right]+d\left[\hat\xi\cdot\delta\hat{\boldsymbol\Theta}^{EM}(\hat\phi,\delta\hat\phi)-\hat\xi\cdot\delta^2\hat{\boldsymbol B}^{EM}_{\mathcal I^+}(\hat\phi)\right]\, .
\label{app:exactform_infinity_split}
\end{equation}
The corresponding gravitational boundary contribution on $\mathcal I^+$ has been derived in~\cite{Fu:2026itf}. In the present notation, it gives
\begin{equation}
  d\left[ \hat\xi\cdot\delta\hat{\boldsymbol\Theta}^{GR}(\hat g,\delta\hat g) - \hat\xi\cdot\delta^2\hat{\boldsymbol B}^{GR}_{\mathcal I^+}(\hat g) \right] \overset{\mathcal I^+}{=} \frac{1}{32\pi G}\hat{\boldsymbol\epsilon}^{(3)} \hat \xi^c \partial_c(\hat\tau^{ab}\delta\hat N_{ab})\, .
\label{app:boundary_term_null_infinity_GR}
\end{equation}
For the electromagnetic part, the unphysical symplectic potential is
\begin{equation}
  \hat{\boldsymbol\Theta}^{EM}(\hat\phi,\delta\hat\phi)=-\frac{1}{4\pi G}\hat{\boldsymbol\epsilon}^{(4)}_a\hat F^{ab}\delta\hat A_b\, .
\label{app:theta_EM_infinity}
\end{equation}
A similar reasoning applies on future null infinity. In the absence of background electromagnetic radiation, $\hat\gamma^b{}_c\hat n_a\hat F^{ac}\overset{\mathcal I^+}{=}0$~\cite{Bonga:2019bim}, so $\hat n_a\hat F^{ab}$ is proportional to $\hat n^b$. Together with Eq.~\eqref{app:infinity_maxwell_gauge}, this gives $\underline{\hat{\boldsymbol\Theta}}^{EM}(\hat\phi,\delta\hat\phi)\overset{\mathcal I^+}{=}0$. For general perturbations, we adopt the standard Maxwell symplectic potential on $\mathcal I^+$~\cite{Bonga:2019bim},
\begin{equation}
 \hat{\boldsymbol\Theta}'{}^{EM}(\hat\phi,\delta\hat\phi) \overset{\mathcal I^+}{=} \underline{\hat{\boldsymbol\Theta}}^{EM}(\hat\phi,\delta\hat\phi)\, . 
\label{app:em_boundary_potential_infinity}
\end{equation}
This choice requires no additional electromagnetic boundary correction, so we take $\hat{\boldsymbol B}^{EM}_{\mathcal I^+}(\hat\phi)=0$ through second order. Hence $\delta^2\hat{\boldsymbol B}^{EM}_{\mathcal I^+}(\hat\phi)=0$, while $\delta\underline{\hat{\boldsymbol\Theta}}^{EM}(\hat\phi,\delta\hat\phi)$ remains nonzero for electromagnetic radiation. The electromagnetic exact form becomes
\begin{equation}
  d\left[\hat\xi\cdot\delta\hat{\boldsymbol\Theta}^{EM}(\hat\phi,\delta\hat\phi)-\hat\xi\cdot\delta^2\hat{\boldsymbol B}^{EM}_{\mathcal I^+}(\hat\phi)\right]\overset{\mathcal I^+}{=}\frac{1}{4\pi G}d\left[(\hat t^c\hat\nabla_c\hat u)\hat{\boldsymbol\epsilon}^{(2)}\hat n_a\delta\hat A_b\delta\hat F^{ab}\right]\, .
\label{app:exactform_infinity_EM}
\end{equation}
Thus, after integration over $\mathcal I^+_{12}$, one obtains
\begin{equation}
  \int_{\mathcal I^+_{12}}d\left[\hat\xi\cdot\delta\hat{\boldsymbol\Theta}(\hat\phi,\delta\hat\phi)-\hat\xi\cdot\delta^2\hat{\boldsymbol B}_{\mathcal I^+}(\hat\phi)\right]=\frac{1}{32\pi G}\Delta\int_{\hat{\mathcal C}}\hat{\boldsymbol\epsilon}^{(2)}(\hat t^c\hat\nabla_c\hat u)\hat\tau^{ab}\delta\hat N_{ab}+\frac{1}{4\pi G}\Delta\int_{\hat{\mathcal C}}\hat{\boldsymbol\epsilon}^{(2)}(\hat t^c\hat\nabla_c\hat u)\hat n_a\delta\hat A_b\delta\hat F^{ab}\, .
\label{app:infinity_exact_form_cancellation}
\end{equation}
These two terms cancel the gravitational and electromagnetic boundary contributions on $\mathcal I^+$, respectively.

It follows that the modified canonical energy forms can be defined as
\begin{equation}
  \boldsymbol e_\Phi\overset{\mathcal H^+}{=}\boldsymbol\omega'(\phi;\delta\phi,\mathcal L_\xi\delta\phi)+d\left[\xi\cdot\delta\boldsymbol\Theta(\phi,\delta\phi)-\xi\cdot\delta^2\boldsymbol B_{\mathcal H^+}(\phi)\right]\, ,
\label{app:ePhi_horizon_definition}
\end{equation}
and
\begin{equation}
  \hat{\boldsymbol e}_\Phi\overset{\mathcal I^+}{=}\hat{\boldsymbol\omega}'(\hat\phi;\delta\hat\phi,\mathcal L_{\hat\xi}\delta\hat\phi)+d\left[\hat\xi\cdot\delta\hat{\boldsymbol\Theta}(\hat\phi,\delta\hat\phi)-\hat\xi\cdot\delta^2\hat{\boldsymbol B}_{\mathcal I^+}(\hat\phi)\right]\, .
\label{app:ePhi_infinity_definition}
\end{equation}
Equivalently, the same boundary contributions can be absorbed into the definition of $\mathcal E'[\Sigma(t)]$ in Eq.~\eqref{eq:modified_canonical_energy_prime_definition}. Therefore, Stokes' theorem gives
\begin{equation}
  \mathcal E'[\Sigma(t_2)]-\mathcal E'[\Sigma(t_1)]=-\int_{\mathcal H^+_{12}}\boldsymbol e_\Phi-\int_{\mathcal I^+_{12}}\hat{\boldsymbol e}_\Phi\, .
\label{app:ePhi_balance_law}
\end{equation}
Finally, because the second and third terms in Eq.~\eqref{eq:main_modified_current_expanded} vanish after pullback to $\mathcal H^+$ and $\mathcal I^+$, respectively, Eqs.~\eqref{app:ePhi_horizon_definition} and~\eqref{app:ePhi_infinity_definition} become
\begin{equation}
  \boldsymbol e_\Phi \overset{\mathcal H^+}{=} d\left[ \delta_\phi^2\boldsymbol Q_\xi(\phi) - \xi\cdot\delta^2\boldsymbol B_{\mathcal H^+}(\phi) \right]\, ,
\label{app:ephi_exact_horizon}
\end{equation}
and
\begin{equation}
  \hat{\boldsymbol e}_\Phi \overset{\mathcal I^+}{=} d\left[ \delta_\phi^2\hat{\boldsymbol Q}_{\hat\xi}(\hat\phi) - \hat\xi\cdot\delta^2\hat{\boldsymbol B}_{\mathcal I^+}(\hat\phi) \right]\, .
\label{app:ephi_exact_infinity}
\end{equation}
These relations are used in the derivation of the balance law.

\section{Construction of the second-order entropy formula}
\label{app:entropy}

\subsection{Balance law}
\label{app:entropy_balance_law}

The entropy $(n-2)$-form is defined by~\cite{Hollands:2024vbe}
\begin{equation}
  \boldsymbol S = \frac{2\pi}{\kappa_3} \left[ \boldsymbol Q_\xi(\phi) - \xi\cdot\boldsymbol B_{\mathcal H^+}(\phi) \right]\, ,
\label{app:entropy_form}
\end{equation}
Equivalently, the corresponding $(n-1)$-form on $\mathcal H^+$ is
\begin{equation}
  d\boldsymbol S = 2\pi d\left[ \frac{\boldsymbol Q_\xi(\phi)}{\kappa_3} - \frac{\xi\cdot\boldsymbol B_{\mathcal H^+}(\phi)}{\kappa_3} \right]\, .
\label{app:entropy_n_minus_one_form}
\end{equation}
For Einstein--Maxwell theory, the Noether charge is
\begin{equation}
  \boldsymbol Q_\xi(\phi) = -\frac{1}{16\pi G} \boldsymbol\epsilon^{(4)}_{ab}\nabla^a\xi^b - \frac{1}{8\pi G} \boldsymbol\epsilon^{(4)}_{ab}F^{ab}\xi^cA_c\, .
\label{app:noether_charge_EM}
\end{equation}
On $\mathcal H^+$, using the binormal~\cite{Iyer:1994ys} $\boldsymbol\epsilon_{ab}=k_al_b-l_ak_b$, one has $\boldsymbol\epsilon^{ab}\nabla_a\xi_b=-2\kappa_3$. Thus the Noether charge restricted to a horizon boundary cross section is
\begin{equation}
  \boldsymbol Q_\xi(\phi) = \frac{\kappa_3}{8\pi G}\boldsymbol\epsilon^{(2)} - \frac{1}{8\pi G} \boldsymbol\epsilon^{(4)}_{ab}F^{ab}\xi^cA_c\, .
\label{app:noether_charge_horizon}
\end{equation}
Taking into account Eq.~\eqref{app:horizon_EM_gauge_condition}, the electromagnetic term does not contribute to the entropy variation. Therefore, only the gravitational part of the Noether charge is relevant. Moreover, taking into account $\boldsymbol B^{EM}_{\mathcal H^+}(\phi)=0$ and $\delta^2\boldsymbol B^{EM}_{\mathcal H^+}(\phi)=0$, we have~\cite{Hollands:2024vbe,Fu:2026itf}
\begin{equation}
  \boldsymbol B_{\mathcal H^+}(\phi) = \boldsymbol B_{\mathcal H^+}(g) = \frac{1}{8\pi G}\, \boldsymbol\epsilon^{(3)}\theta_v\, .
  \label{app:BH_GR_definition}
\end{equation}

The variations of surface gravities is encoded in~\cite{Visser:2024pwz} $\delta\kappa_2 \overset{\mathcal H^+}{=} -k^a\nabla_a(l_b\delta\xi^b)$ and $\delta\kappa_3 \overset{\mathcal H^+}{=} l_{[a}k_{b]}\nabla^a\delta\xi^b$. Since $\delta\xi^a$ is proportional to $k^a$ on $\mathcal H^+$, we write
\begin{equation}
\delta\xi^a \overset{\mathcal H^+}{=} \delta h\,k^a\, , \qquad \delta h(v,x)=\int_0^v\delta\kappa_2(v',x)\,dv'\, .
\label{app:delta_h_definition}
\end{equation}
We then take the second variation of Eq.~\eqref{app:entropy_n_minus_one_form}. For the first term, using Eq.~\eqref{app:horizon_EM_gauge_condition} and~\eqref{app:noether_charge_horizon}, one finds
\begin{equation}
  \delta^2\left[ \frac{\boldsymbol Q_\xi(\phi)}{\kappa_3} \right] = \frac{1}{8\pi G}\delta_\phi^2\boldsymbol\epsilon^{(2)} = \frac{1}{\kappa} \delta_\phi^2\boldsymbol Q_\xi(g)\, .
\label{app:second_variation_noether_over_kappa}
\end{equation}
For the second term, using Eq.~\eqref{app:BH_GR_definition}, one obtains
\begin{equation}
  \delta^2\left[ \frac{\xi\cdot\boldsymbol B_{\mathcal H^+}(\phi)}{\kappa_3} \right] \overset{\mathcal H^+}{=} \frac{1}{\kappa} \xi\cdot\delta_\phi^2\boldsymbol B_{\mathcal H^+}(g) + \frac{2}{\kappa} \left[ \delta\xi\cdot\delta\boldsymbol B_{\mathcal H^+}(g) - \frac{\delta\kappa_3}{\kappa} \xi\cdot\delta\boldsymbol B_{\mathcal H^+}(g) \right]\, .
\label{app:second_variation_boundary_over_kappa}
\end{equation}
Combining Eqs.~\eqref{app:entropy_n_minus_one_form}, \eqref{app:second_variation_noether_over_kappa}, and \eqref{app:second_variation_boundary_over_kappa}, we obtain
\begin{equation}
  \frac{\kappa}{2\pi}d\delta^2\boldsymbol S \overset{\mathcal H^+}{=} d\left[ \delta_\phi^2\boldsymbol Q_\xi(g) - \xi\cdot\delta_\phi^2\boldsymbol B_{\mathcal H^+}(g) \right] - 2d\left[ \delta\xi\cdot\delta\boldsymbol B_{\mathcal H^+}(g) - \frac{\delta\kappa_3}{\kappa} \xi\cdot\delta\boldsymbol B_{\mathcal H^+}(g) \right]\, .
\label{app:second_order_entropy_form_horizon}
\end{equation}
Employing Eq.~\eqref{app:ephi_exact_horizon} gives
\begin{equation}
  \frac{\kappa}{2\pi}\Delta\delta^2S = \int_{\mathcal H^+_{12}}\boldsymbol e_\Phi - \int_{\mathcal H^+_{12}}2d\left[ \delta\xi\cdot\delta\boldsymbol B_{\mathcal H^+}(g) - \frac{\delta\kappa_3}{\kappa} \xi\cdot\delta\boldsymbol B_{\mathcal H^+}(g) \right]\, .
\label{app:entropy_balance_start}
\end{equation}
Using Eqs.~\eqref{app:BH_GR_definition} and \eqref{app:delta_h_definition}, the correction term associated with the variation of $\xi^a$ can be evaluated explicitly. Therefore,
\begin{equation}
  \frac{\kappa}{2\pi}\Delta\delta^2S = \int_{\mathcal H^+_{12}}\boldsymbol e_\Phi + \frac{1}{4\pi G} \Delta\int_{\mathcal C} dA_\epsilon \left( v\delta\kappa_3-\delta h \right)\delta\theta_v\, .
\label{app:delta_xi_boundary_contribution}
\end{equation}

It remains to reduce the first term. This step follows the same geometric reduction as the corresponding gravitational calculation in~\cite{Fu:2026itf}: using the first- and second-order Raychaudhuri equations, the shear contraction identity, and integration by parts along the horizon generators, the modified canonical energy contribution can be rewritten entirely as a difference between the two boundary cross sections. Therefore, applying the result of~\cite{Fu:2026itf} to Eq.~\eqref{app:delta_xi_boundary_contribution}, one obtains
\begin{equation}
  \frac{\kappa}{2\pi}\Delta\delta^2S = \frac{\kappa}{8\pi G} \Delta\delta^2 \int_{\mathcal C} dA_\epsilon(1-v\theta_v) + \frac{1}{4\pi G} \Delta\int_{\mathcal C} dA_\epsilon \left( v\delta\kappa_3-\delta h \right)\delta\theta_v\, .
\label{app:canonical_energy_simplified}
\end{equation}
Since this relation holds for arbitrary boundary cross sections, we arrive at
\begin{equation}
  \delta^2S = \frac{1}{4G} \delta^2\int_{\mathcal C}dA_\epsilon(1-v\theta_v) + \frac{1}{2G\kappa} \int_{\mathcal C}dA_\epsilon \left( v\delta\kappa_3-\delta h \right)\delta\theta_v\, .
\label{app:entropy_reduced_formula}
\end{equation}
This reproduces Eq.~\eqref{eq:main_entropy_formula_expanded}.

\bibliography{mainRef}

\begin{thebibliography}{33}%
\makeatletter
\providecommand \@ifxundefined [1]{%
 \@ifx{#1\undefined}
}%
\providecommand \@ifnum [1]{%
 \ifnum #1\expandafter \@firstoftwo
 \else \expandafter \@secondoftwo
 \fi
}%
\providecommand \@ifx [1]{%
 \ifx #1\expandafter \@firstoftwo
 \else \expandafter \@secondoftwo
 \fi
}%
\providecommand \natexlab [1]{#1}%
\providecommand \enquote  [1]{``#1''}%
\providecommand \bibnamefont  [1]{#1}%
\providecommand \bibfnamefont [1]{#1}%
\providecommand \citenamefont [1]{#1}%
\providecommand \href@noop [0]{\@secondoftwo}%
\providecommand \href [0]{\begingroup \@sanitize@url \@href}%
\providecommand \@href[1]{\@@startlink{#1}\@@href}%
\providecommand \@@href[1]{\endgroup#1\@@endlink}%
\providecommand \@sanitize@url [0]{\catcode `\\12\catcode `\$12\catcode
  `\&12\catcode `\#12\catcode `\^12\catcode `\_12\catcode `\%12\relax}%
\providecommand \@@startlink[1]{}%
\providecommand \@@endlink[0]{}%
\providecommand \url  [0]{\begingroup\@sanitize@url \@url }%
\providecommand \@url [1]{\endgroup\@href {#1}{\urlprefix }}%
\providecommand \urlprefix  [0]{URL }%
\providecommand \Eprint [0]{\href }%
\providecommand \doibase [0]{http://dx.doi.org/}%
\providecommand \selectlanguage [0]{\@gobble}%
\providecommand \bibinfo  [0]{\@secondoftwo}%
\providecommand \bibfield  [0]{\@secondoftwo}%
\providecommand \translation [1]{[#1]}%
\providecommand \BibitemOpen [0]{}%
\providecommand \bibitemStop [0]{}%
\providecommand \bibitemNoStop [0]{.\EOS\space}%
\providecommand \EOS [0]{\spacefactor3000\relax}%
\providecommand \BibitemShut  [1]{\csname bibitem#1\endcsname}%
\let\auto@bib@innerbib\@empty
\bibitem [{\citenamefont {Bekenstein}(1973)}]{Bekenstein:1973ur}%
  \BibitemOpen
  \bibfield  {author} {\bibinfo {author} {\bibfnamefont {J.~D.}\ \bibnamefont
  {Bekenstein}},\ }\href {\doibase 10.1103/PhysRevD.7.2333} {\bibfield
  {journal} {\bibinfo  {journal} {Phys. Rev. D}\ }\textbf {\bibinfo {volume}
  {7}},\ \bibinfo {pages} {2333} (\bibinfo {year} {1973})}\BibitemShut
  {NoStop}%
\bibitem [{\citenamefont {Bardeen}\ \emph {et~al.}(1973)\citenamefont
  {Bardeen}, \citenamefont {Carter},\ and\ \citenamefont
  {Hawking}}]{Bardeen:1973gs}%
  \BibitemOpen
  \bibfield  {author} {\bibinfo {author} {\bibfnamefont {J.~M.}\ \bibnamefont
  {Bardeen}}, \bibinfo {author} {\bibfnamefont {B.}~\bibnamefont {Carter}}, \
  and\ \bibinfo {author} {\bibfnamefont {S.~W.}\ \bibnamefont {Hawking}},\
  }\href {\doibase 10.1007/BF01645742} {\bibfield  {journal} {\bibinfo
  {journal} {Commun. Math. Phys.}\ }\textbf {\bibinfo {volume} {31}},\ \bibinfo
  {pages} {161} (\bibinfo {year} {1973})}\BibitemShut {NoStop}%
\bibitem [{\citenamefont {Hawking}(1975)}]{Hawking:1975vcx}%
  \BibitemOpen
  \bibfield  {author} {\bibinfo {author} {\bibfnamefont {S.~W.}\ \bibnamefont
  {Hawking}},\ }\href {\doibase 10.1007/BF02345020} {\bibfield  {journal}
  {\bibinfo  {journal} {Commun. Math. Phys.}\ }\textbf {\bibinfo {volume}
  {43}},\ \bibinfo {pages} {199} (\bibinfo {year} {1975})},\ \bibinfo {note}
  {[Erratum: Commun.Math.Phys. 46, 206 (1976)]}\BibitemShut {NoStop}%
\bibitem [{\citenamefont {Wald}(2001)}]{Wald:1999vt}%
  \BibitemOpen
  \bibfield  {author} {\bibinfo {author} {\bibfnamefont {R.~M.}\ \bibnamefont
  {Wald}},\ }\href {\doibase 10.12942/lrr-2001-6} {\bibfield  {journal}
  {\bibinfo  {journal} {Living Rev. Rel.}\ }\textbf {\bibinfo {volume} {4}},\
  \bibinfo {pages} {6} (\bibinfo {year} {2001})},\ \Eprint
  {http://arxiv.org/abs/gr-qc/9912119} {arXiv:gr-qc/9912119} \BibitemShut
  {NoStop}%
\bibitem [{\citenamefont {Lee}\ and\ \citenamefont {Wald}(1990)}]{Lee:1990nz}%
  \BibitemOpen
  \bibfield  {author} {\bibinfo {author} {\bibfnamefont {J.}~\bibnamefont
  {Lee}}\ and\ \bibinfo {author} {\bibfnamefont {R.~M.}\ \bibnamefont {Wald}},\
  }\href {\doibase 10.1063/1.528801} {\bibfield  {journal} {\bibinfo  {journal}
  {J. Math. Phys.}\ }\textbf {\bibinfo {volume} {31}},\ \bibinfo {pages} {725}
  (\bibinfo {year} {1990})}\BibitemShut {NoStop}%
\bibitem [{\citenamefont {Wald}(1990)}]{Wald:1990mme}%
  \BibitemOpen
  \bibfield  {author} {\bibinfo {author} {\bibfnamefont {R.~M.}\ \bibnamefont
  {Wald}},\ }\href {\doibase 10.1063/1.528839} {\bibfield  {journal} {\bibinfo
  {journal} {J. Math. Phys.}\ }\textbf {\bibinfo {volume} {31}},\ \bibinfo
  {pages} {2378} (\bibinfo {year} {1990})}\BibitemShut {NoStop}%
\bibitem [{\citenamefont {Wald}(1993)}]{Wald:1993nt}%
  \BibitemOpen
  \bibfield  {author} {\bibinfo {author} {\bibfnamefont {R.~M.}\ \bibnamefont
  {Wald}},\ }\href {\doibase 10.1103/PhysRevD.48.R3427} {\bibfield  {journal}
  {\bibinfo  {journal} {Phys. Rev. D}\ }\textbf {\bibinfo {volume} {48}},\
  \bibinfo {pages} {R3427} (\bibinfo {year} {1993})},\ \Eprint
  {http://arxiv.org/abs/gr-qc/9307038} {arXiv:gr-qc/9307038} \BibitemShut
  {NoStop}%
\bibitem [{\citenamefont {Jacobson}\ \emph {et~al.}(1994)\citenamefont
  {Jacobson}, \citenamefont {Kang},\ and\ \citenamefont
  {Myers}}]{Jacobson:1993vj}%
  \BibitemOpen
  \bibfield  {author} {\bibinfo {author} {\bibfnamefont {T.}~\bibnamefont
  {Jacobson}}, \bibinfo {author} {\bibfnamefont {G.}~\bibnamefont {Kang}}, \
  and\ \bibinfo {author} {\bibfnamefont {R.~C.}\ \bibnamefont {Myers}},\ }\href
  {\doibase 10.1103/PhysRevD.49.6587} {\bibfield  {journal} {\bibinfo
  {journal} {Phys. Rev. D}\ }\textbf {\bibinfo {volume} {49}},\ \bibinfo
  {pages} {6587} (\bibinfo {year} {1994})},\ \Eprint
  {http://arxiv.org/abs/gr-qc/9312023} {arXiv:gr-qc/9312023} \BibitemShut
  {NoStop}%
\bibitem [{\citenamefont {Iyer}\ and\ \citenamefont
  {Wald}(1994)}]{Iyer:1994ys}%
  \BibitemOpen
  \bibfield  {author} {\bibinfo {author} {\bibfnamefont {V.}~\bibnamefont
  {Iyer}}\ and\ \bibinfo {author} {\bibfnamefont {R.~M.}\ \bibnamefont
  {Wald}},\ }\href {\doibase 10.1103/PhysRevD.50.846} {\bibfield  {journal}
  {\bibinfo  {journal} {Phys. Rev. D}\ }\textbf {\bibinfo {volume} {50}},\
  \bibinfo {pages} {846} (\bibinfo {year} {1994})},\ \Eprint
  {http://arxiv.org/abs/gr-qc/9403028} {arXiv:gr-qc/9403028} \BibitemShut
  {NoStop}%
\bibitem [{\citenamefont {Ashtekar}\ and\ \citenamefont
  {Krishnan}(2004)}]{Ashtekar_2004}%
  \BibitemOpen
  \bibfield  {author} {\bibinfo {author} {\bibfnamefont {A.}~\bibnamefont
  {Ashtekar}}\ and\ \bibinfo {author} {\bibfnamefont {B.}~\bibnamefont
  {Krishnan}},\ }\href {\doibase 10.12942/lrr-2004-10} {\bibfield  {journal}
  {\bibinfo  {journal} {Living Reviews in Relativity}\ }\textbf {\bibinfo
  {volume} {7}} (\bibinfo {year} {2004}),\ 10.12942/lrr-2004-10}\BibitemShut
  {NoStop}%
\bibitem [{\citenamefont {Hollands}\ \emph {et~al.}(2024)\citenamefont
  {Hollands}, \citenamefont {Wald},\ and\ \citenamefont
  {Zhang}}]{Hollands:2024vbe}%
  \BibitemOpen
  \bibfield  {author} {\bibinfo {author} {\bibfnamefont {S.}~\bibnamefont
  {Hollands}}, \bibinfo {author} {\bibfnamefont {R.~M.}\ \bibnamefont {Wald}},
  \ and\ \bibinfo {author} {\bibfnamefont {V.~G.}\ \bibnamefont {Zhang}},\
  }\href {\doibase 10.1103/PhysRevD.110.024070} {\bibfield  {journal} {\bibinfo
   {journal} {Phys. Rev. D}\ }\textbf {\bibinfo {volume} {110}},\ \bibinfo
  {pages} {024070} (\bibinfo {year} {2024})},\ \Eprint
  {http://arxiv.org/abs/2402.00818} {arXiv:2402.00818 [hep-th]} \BibitemShut
  {NoStop}%
\bibitem [{\citenamefont {Dong}(2014)}]{Dong:2013qoa}%
  \BibitemOpen
  \bibfield  {author} {\bibinfo {author} {\bibfnamefont {X.}~\bibnamefont
  {Dong}},\ }\href {\doibase 10.1007/JHEP01(2014)044} {\bibfield  {journal}
  {\bibinfo  {journal} {JHEP}\ }\textbf {\bibinfo {volume} {01}},\ \bibinfo
  {pages} {044} (\bibinfo {year} {2014})},\ \Eprint
  {http://arxiv.org/abs/1310.5713} {arXiv:1310.5713 [hep-th]} \BibitemShut
  {NoStop}%
\bibitem [{\citenamefont {Wall}(2015)}]{Wall:2015raa}%
  \BibitemOpen
  \bibfield  {author} {\bibinfo {author} {\bibfnamefont {A.~C.}\ \bibnamefont
  {Wall}},\ }\href {\doibase 10.1142/S0218271815440149} {\bibfield  {journal}
  {\bibinfo  {journal} {Int. J. Mod. Phys. D}\ }\textbf {\bibinfo {volume}
  {24}},\ \bibinfo {pages} {1544014} (\bibinfo {year} {2015})},\ \Eprint
  {http://arxiv.org/abs/1504.08040} {arXiv:1504.08040 [gr-qc]} \BibitemShut
  {NoStop}%
\bibitem [{\citenamefont {Rignon-Bret}(2023)}]{Rignon-Bret:2023fjq}%
  \BibitemOpen
  \bibfield  {author} {\bibinfo {author} {\bibfnamefont {A.}~\bibnamefont
  {Rignon-Bret}},\ }\href {\doibase 10.1103/PhysRevD.108.044069} {\bibfield
  {journal} {\bibinfo  {journal} {Phys. Rev. D}\ }\textbf {\bibinfo {volume}
  {108}},\ \bibinfo {pages} {044069} (\bibinfo {year} {2023})},\ \Eprint
  {http://arxiv.org/abs/2303.07262} {arXiv:2303.07262 [gr-qc]} \BibitemShut
  {NoStop}%
\bibitem [{\citenamefont {Furugori}\ \emph {et~al.}(2025)\citenamefont
  {Furugori}, \citenamefont {Nishii}, \citenamefont {Yoshida},\ and\
  \citenamefont {Yoshimura}}]{Furugori:2025pmn}%
  \BibitemOpen
  \bibfield  {author} {\bibinfo {author} {\bibfnamefont {H.}~\bibnamefont
  {Furugori}}, \bibinfo {author} {\bibfnamefont {K.}~\bibnamefont {Nishii}},
  \bibinfo {author} {\bibfnamefont {D.}~\bibnamefont {Yoshida}}, \ and\
  \bibinfo {author} {\bibfnamefont {K.}~\bibnamefont {Yoshimura}},\ }\href
  {\doibase 10.1103/f2t6-7t61} {\bibfield  {journal} {\bibinfo  {journal}
  {Phys. Rev. D}\ }\textbf {\bibinfo {volume} {112}},\ \bibinfo {pages}
  {104049} (\bibinfo {year} {2025})},\ \Eprint
  {http://arxiv.org/abs/2507.14105} {arXiv:2507.14105 [gr-qc]} \BibitemShut
  {NoStop}%
\bibitem [{\citenamefont {Kong}\ \emph {et~al.}(2025)\citenamefont {Kong},
  \citenamefont {Tian}, \citenamefont {Zhang},\ and\ \citenamefont
  {Zhao}}]{Kong:2024sqc}%
  \BibitemOpen
  \bibfield  {author} {\bibinfo {author} {\bibfnamefont {D.}~\bibnamefont
  {Kong}}, \bibinfo {author} {\bibfnamefont {Y.}~\bibnamefont {Tian}}, \bibinfo
  {author} {\bibfnamefont {H.}~\bibnamefont {Zhang}}, \ and\ \bibinfo {author}
  {\bibfnamefont {J.}~\bibnamefont {Zhao}},\ }\href {\doibase
  10.1103/PhysRevD.111.084005} {\bibfield  {journal} {\bibinfo  {journal}
  {Phys. Rev. D}\ }\textbf {\bibinfo {volume} {111}},\ \bibinfo {pages}
  {084005} (\bibinfo {year} {2025})},\ \Eprint
  {http://arxiv.org/abs/2412.00647} {arXiv:2412.00647 [hep-th]} \BibitemShut
  {NoStop}%
\bibitem [{\citenamefont {Fu}\ \emph {et~al.}(2026)\citenamefont {Fu},
  \citenamefont {Ji}, \citenamefont {Zhou},\ and\ \citenamefont
  {Cao}}]{Fu:2026itf}%
  \BibitemOpen
  \bibfield  {author} {\bibinfo {author} {\bibfnamefont {W.-T.}\ \bibnamefont
  {Fu}}, \bibinfo {author} {\bibfnamefont {M.-F.}\ \bibnamefont {Ji}}, \bibinfo
  {author} {\bibfnamefont {Y.-S.}\ \bibnamefont {Zhou}}, \ and\ \bibinfo
  {author} {\bibfnamefont {L.-M.}\ \bibnamefont {Cao}},\ }\href@noop {} {\
  (\bibinfo {year} {2026})},\ \Eprint {http://arxiv.org/abs/2606.16757}
  {arXiv:2606.16757 [gr-qc]} \BibitemShut {NoStop}%
\bibitem [{\citenamefont {Ashtekar}\ \emph {et~al.}(2025)\citenamefont
  {Ashtekar}, \citenamefont {Paraizo},\ and\ \citenamefont
  {Shu}}]{Ashtekar:2025qqa}%
  \BibitemOpen
  \bibfield  {author} {\bibinfo {author} {\bibfnamefont {A.}~\bibnamefont
  {Ashtekar}}, \bibinfo {author} {\bibfnamefont {D.~E.}\ \bibnamefont
  {Paraizo}}, \ and\ \bibinfo {author} {\bibfnamefont {J.}~\bibnamefont
  {Shu}},\ }\href@noop {} {\  (\bibinfo {year} {2025})},\ \Eprint
  {http://arxiv.org/abs/2512.11659} {arXiv:2512.11659 [gr-qc]} \BibitemShut
  {NoStop}%
\bibitem [{\citenamefont {Ashtekar}\ \emph {et~al.}(2026)\citenamefont
  {Ashtekar}, \citenamefont {Paraizo},\ and\ \citenamefont
  {Shu}}]{Ashtekar:2026jdz}%
  \BibitemOpen
  \bibfield  {author} {\bibinfo {author} {\bibfnamefont {A.}~\bibnamefont
  {Ashtekar}}, \bibinfo {author} {\bibfnamefont {D.~E.}\ \bibnamefont
  {Paraizo}}, \ and\ \bibinfo {author} {\bibfnamefont {J.}~\bibnamefont
  {Shu}},\ }\href@noop {} {\  (\bibinfo {year} {2026})},\ \Eprint
  {http://arxiv.org/abs/2604.00170} {arXiv:2604.00170 [gr-qc]} \BibitemShut
  {NoStop}%
\bibitem [{\citenamefont {Chrusciel}\ \emph {et~al.}(2012)\citenamefont
  {Chrusciel}, \citenamefont {Lopes~Costa},\ and\ \citenamefont
  {Heusler}}]{Chrusciel:2012jk}%
  \BibitemOpen
  \bibfield  {author} {\bibinfo {author} {\bibfnamefont {P.~T.}\ \bibnamefont
  {Chrusciel}}, \bibinfo {author} {\bibfnamefont {J.}~\bibnamefont
  {Lopes~Costa}}, \ and\ \bibinfo {author} {\bibfnamefont {M.}~\bibnamefont
  {Heusler}},\ }\href {\doibase 10.12942/lrr-2012-7} {\bibfield  {journal}
  {\bibinfo  {journal} {Living Rev. Rel.}\ }\textbf {\bibinfo {volume} {15}},\
  \bibinfo {pages} {7} (\bibinfo {year} {2012})},\ \Eprint
  {http://arxiv.org/abs/1205.6112} {arXiv:1205.6112 [gr-qc]} \BibitemShut
  {NoStop}%
\bibitem [{\citenamefont {Gao}(2003)}]{Gao:2003ys}%
  \BibitemOpen
  \bibfield  {author} {\bibinfo {author} {\bibfnamefont {S.}~\bibnamefont
  {Gao}},\ }\href {\doibase 10.1103/PhysRevD.68.044016} {\bibfield  {journal}
  {\bibinfo  {journal} {Phys. Rev. D}\ }\textbf {\bibinfo {volume} {68}},\
  \bibinfo {pages} {044016} (\bibinfo {year} {2003})},\ \Eprint
  {http://arxiv.org/abs/gr-qc/0304094} {arXiv:gr-qc/0304094} \BibitemShut
  {NoStop}%
\bibitem [{\citenamefont {Visser}\ and\ \citenamefont
  {Yan}(2026)}]{Visser:2025jnf}%
  \BibitemOpen
  \bibfield  {author} {\bibinfo {author} {\bibfnamefont {M.~R.}\ \bibnamefont
  {Visser}}\ and\ \bibinfo {author} {\bibfnamefont {Z.}~\bibnamefont {Yan}},\
  }\href {\doibase 10.1007/JHEP02(2026)003} {\bibfield  {journal} {\bibinfo
  {journal} {JHEP}\ }\textbf {\bibinfo {volume} {02}},\ \bibinfo {pages} {003}
  (\bibinfo {year} {2026})},\ \Eprint {http://arxiv.org/abs/2510.20747}
  {arXiv:2510.20747 [hep-th]} \BibitemShut {NoStop}%
\bibitem [{\citenamefont {Visser}\ and\ \citenamefont
  {Yan}(2024)}]{Visser:2024pwz}%
  \BibitemOpen
  \bibfield  {author} {\bibinfo {author} {\bibfnamefont {M.~R.}\ \bibnamefont
  {Visser}}\ and\ \bibinfo {author} {\bibfnamefont {Z.}~\bibnamefont {Yan}},\
  }\href {\doibase 10.1007/JHEP10(2024)029} {\bibfield  {journal} {\bibinfo
  {journal} {JHEP}\ }\textbf {\bibinfo {volume} {10}},\ \bibinfo {pages} {029}
  (\bibinfo {year} {2024})},\ \Eprint {http://arxiv.org/abs/2403.07140}
  {arXiv:2403.07140 [hep-th]} \BibitemShut {NoStop}%
\bibitem [{\citenamefont {Jia}\ \emph {et~al.}(2026)\citenamefont {Jia},
  \citenamefont {Qi},\ and\ \citenamefont {Gao}}]{Jia:2025tgf}%
  \BibitemOpen
  \bibfield  {author} {\bibinfo {author} {\bibfnamefont {W.}~\bibnamefont
  {Jia}}, \bibinfo {author} {\bibfnamefont {Q.}~\bibnamefont {Qi}}, \ and\
  \bibinfo {author} {\bibfnamefont {C.}~\bibnamefont {Gao}},\ }\href {\doibase
  10.1103/8tl5-1qll} {\bibfield  {journal} {\bibinfo  {journal} {Phys. Rev. D}\
  }\textbf {\bibinfo {volume} {113}},\ \bibinfo {pages} {104040} (\bibinfo
  {year} {2026})},\ \Eprint {http://arxiv.org/abs/2509.05700} {arXiv:2509.05700
  [hep-th]} \BibitemShut {NoStop}%
\bibitem [{\citenamefont {Hollands}\ \emph {et~al.}(2022)\citenamefont
  {Hollands}, \citenamefont {Kov{\'a}cs},\ and\ \citenamefont
  {Reall}}]{Hollands:2022fkn}%
  \BibitemOpen
  \bibfield  {author} {\bibinfo {author} {\bibfnamefont {S.}~\bibnamefont
  {Hollands}}, \bibinfo {author} {\bibfnamefont {{\'A}.~D.}\ \bibnamefont
  {Kov{\'a}cs}}, \ and\ \bibinfo {author} {\bibfnamefont {H.~S.}\ \bibnamefont
  {Reall}},\ }\href {\doibase 10.1007/JHEP08(2022)258} {\bibfield  {journal}
  {\bibinfo  {journal} {JHEP}\ }\textbf {\bibinfo {volume} {08}},\ \bibinfo
  {pages} {258} (\bibinfo {year} {2022})},\ \Eprint
  {http://arxiv.org/abs/2205.15341} {arXiv:2205.15341 [hep-th]} \BibitemShut
  {NoStop}%
\bibitem [{\citenamefont {Bhattacharyya}\ \emph {et~al.}(2021)\citenamefont
  {Bhattacharyya}, \citenamefont {Dhivakar}, \citenamefont {Dinda},
  \citenamefont {Kundu}, \citenamefont {Patra},\ and\ \citenamefont
  {Roy}}]{Bhattacharyya:2021jhr}%
  \BibitemOpen
  \bibfield  {author} {\bibinfo {author} {\bibfnamefont {S.}~\bibnamefont
  {Bhattacharyya}}, \bibinfo {author} {\bibfnamefont {P.}~\bibnamefont
  {Dhivakar}}, \bibinfo {author} {\bibfnamefont {A.}~\bibnamefont {Dinda}},
  \bibinfo {author} {\bibfnamefont {N.}~\bibnamefont {Kundu}}, \bibinfo
  {author} {\bibfnamefont {M.}~\bibnamefont {Patra}}, \ and\ \bibinfo {author}
  {\bibfnamefont {S.}~\bibnamefont {Roy}},\ }\href {\doibase
  10.1007/JHEP09(2021)169} {\bibfield  {journal} {\bibinfo  {journal} {JHEP}\
  }\textbf {\bibinfo {volume} {09}},\ \bibinfo {pages} {169} (\bibinfo {year}
  {2021})},\ \Eprint {http://arxiv.org/abs/2105.06455} {arXiv:2105.06455
  [hep-th]} \BibitemShut {NoStop}%
\bibitem [{\citenamefont {Wall}\ and\ \citenamefont
  {Yan}(2024)}]{Wall:2024lbd}%
  \BibitemOpen
  \bibfield  {author} {\bibinfo {author} {\bibfnamefont {A.~C.}\ \bibnamefont
  {Wall}}\ and\ \bibinfo {author} {\bibfnamefont {Z.}~\bibnamefont {Yan}},\
  }\href {\doibase 10.1103/PhysRevD.110.084005} {\bibfield  {journal} {\bibinfo
   {journal} {Phys. Rev. D}\ }\textbf {\bibinfo {volume} {110}},\ \bibinfo
  {pages} {084005} (\bibinfo {year} {2024})},\ \Eprint
  {http://arxiv.org/abs/2402.05411} {arXiv:2402.05411 [gr-qc]} \BibitemShut
  {NoStop}%
\bibitem [{\citenamefont {Sorce}\ and\ \citenamefont
  {Wald}(2017)}]{Sorce:2017dst}%
  \BibitemOpen
  \bibfield  {author} {\bibinfo {author} {\bibfnamefont {J.}~\bibnamefont
  {Sorce}}\ and\ \bibinfo {author} {\bibfnamefont {R.~M.}\ \bibnamefont
  {Wald}},\ }\href {\doibase 10.1103/PhysRevD.96.104014} {\bibfield  {journal}
  {\bibinfo  {journal} {Phys. Rev. D}\ }\textbf {\bibinfo {volume} {96}},\
  \bibinfo {pages} {104014} (\bibinfo {year} {2017})},\ \Eprint
  {http://arxiv.org/abs/1707.05862} {arXiv:1707.05862 [gr-qc]} \BibitemShut
  {NoStop}%
\bibitem [{\citenamefont {Hollands}\ and\ \citenamefont
  {Wald}(2013)}]{Hollands:2012sf}%
  \BibitemOpen
  \bibfield  {author} {\bibinfo {author} {\bibfnamefont {S.}~\bibnamefont
  {Hollands}}\ and\ \bibinfo {author} {\bibfnamefont {R.~M.}\ \bibnamefont
  {Wald}},\ }\href {\doibase 10.1007/s00220-012-1638-1} {\bibfield  {journal}
  {\bibinfo  {journal} {Commun. Math. Phys.}\ }\textbf {\bibinfo {volume}
  {321}},\ \bibinfo {pages} {629} (\bibinfo {year} {2013})},\ \Eprint
  {http://arxiv.org/abs/1201.0463} {arXiv:1201.0463 [gr-qc]} \BibitemShut
  {NoStop}%
\bibitem [{\citenamefont {Wald}\ and\ \citenamefont
  {Zoupas}(2000)}]{Wald:1999wa}%
  \BibitemOpen
  \bibfield  {author} {\bibinfo {author} {\bibfnamefont {R.~M.}\ \bibnamefont
  {Wald}}\ and\ \bibinfo {author} {\bibfnamefont {A.}~\bibnamefont {Zoupas}},\
  }\href {\doibase 10.1103/PhysRevD.61.084027} {\bibfield  {journal} {\bibinfo
  {journal} {Phys. Rev. D}\ }\textbf {\bibinfo {volume} {61}},\ \bibinfo
  {pages} {084027} (\bibinfo {year} {2000})},\ \Eprint
  {http://arxiv.org/abs/gr-qc/9911095} {arXiv:gr-qc/9911095} \BibitemShut
  {NoStop}%
\bibitem [{\citenamefont {Bhattacharyya}\ \emph {et~al.}(2023)\citenamefont
  {Bhattacharyya}, \citenamefont {Jethwani}, \citenamefont {Patra},\ and\
  \citenamefont {Roy}}]{Bhattacharyya:2022njk}%
  \BibitemOpen
  \bibfield  {author} {\bibinfo {author} {\bibfnamefont {S.}~\bibnamefont
  {Bhattacharyya}}, \bibinfo {author} {\bibfnamefont {P.}~\bibnamefont
  {Jethwani}}, \bibinfo {author} {\bibfnamefont {M.}~\bibnamefont {Patra}}, \
  and\ \bibinfo {author} {\bibfnamefont {S.}~\bibnamefont {Roy}},\ }\href
  {\doibase 10.1103/PhysRevD.108.104032} {\bibfield  {journal} {\bibinfo
  {journal} {Phys. Rev. D}\ }\textbf {\bibinfo {volume} {108}},\ \bibinfo
  {pages} {104032} (\bibinfo {year} {2023})},\ \Eprint
  {http://arxiv.org/abs/2204.08447} {arXiv:2204.08447 [hep-th]} \BibitemShut
  {NoStop}%
\bibitem [{\citenamefont {Kar}\ \emph {et~al.}(2024)\citenamefont {Kar},
  \citenamefont {Dhivakar}, \citenamefont {Roy}, \citenamefont {Panda},\ and\
  \citenamefont {Shaikh}}]{Kar:2024dqk}%
  \BibitemOpen
  \bibfield  {author} {\bibinfo {author} {\bibfnamefont {A.}~\bibnamefont
  {Kar}}, \bibinfo {author} {\bibfnamefont {P.}~\bibnamefont {Dhivakar}},
  \bibinfo {author} {\bibfnamefont {S.}~\bibnamefont {Roy}}, \bibinfo {author}
  {\bibfnamefont {B.}~\bibnamefont {Panda}}, \ and\ \bibinfo {author}
  {\bibfnamefont {A.}~\bibnamefont {Shaikh}},\ }\href {\doibase
  10.1007/JHEP07(2024)016} {\bibfield  {journal} {\bibinfo  {journal} {JHEP}\
  }\textbf {\bibinfo {volume} {07}},\ \bibinfo {pages} {016} (\bibinfo {year}
  {2024})},\ \Eprint {http://arxiv.org/abs/2403.04749} {arXiv:2403.04749
  [hep-th]} \BibitemShut {NoStop}%
\bibitem [{\citenamefont {Bonga}\ \emph {et~al.}(2020)\citenamefont {Bonga},
  \citenamefont {Grant},\ and\ \citenamefont {Prabhu}}]{Bonga:2019bim}%
  \BibitemOpen
  \bibfield  {author} {\bibinfo {author} {\bibfnamefont {B.}~\bibnamefont
  {Bonga}}, \bibinfo {author} {\bibfnamefont {A.~M.}\ \bibnamefont {Grant}}, \
  and\ \bibinfo {author} {\bibfnamefont {K.}~\bibnamefont {Prabhu}},\ }\href
  {\doibase 10.1103/PhysRevD.101.044013} {\bibfield  {journal} {\bibinfo
  {journal} {Phys. Rev. D}\ }\textbf {\bibinfo {volume} {101}},\ \bibinfo
  {pages} {044013} (\bibinfo {year} {2020})},\ \Eprint
  {http://arxiv.org/abs/1911.04514} {arXiv:1911.04514 [gr-qc]} \BibitemShut
  {NoStop}%
\end{thebibliography}%
\bibliographystyle{apsrev4-1}

\end{document}